\documentclass[12pt,a4paper]{article}

\usepackage[utf8]{inputenc}
\usepackage{amsmath}
\usepackage{amsfonts}
\usepackage{amssymb}
\usepackage{amsthm}
\usepackage{graphicx}
\usepackage{todonotes}
\usepackage{natbib}
\usepackage{url}
\usepackage[boxruled,vlined,linesnumbered]{algorithm2e}
\usepackage{caption}
\usepackage{subcaption}
\usepackage{lineno}
\usepackage{tcolorbox}
\usepackage{hyperref}
\usepackage{fancyhdr}
\AtBeginDocument{}
\hypersetup{colorlinks=true,linkcolor=black,citecolor=black,filecolor=black,urlcolor=black}

\providecommand{\keywords}[1]{\textbf{\textit{Keywords:}} #1}

\theoremstyle{definition}
\newtheorem{definition}{Definition}

\author{Jonathan D. Rosenblatt \\ 
	Department of IE\&M and \\
	Zlotowsky Center for Neuroscience, \\
	Ben Gurion University of the Negev, Israel. 
	\and Yuval Benjamini\\
	Department of Statistics, \\
	Hebrew University, Israel
	\and Roee Gilron \\ 
	Movement Disorders and Neuromodulation Center,\\
	University of California, San Francisco. 
	\and Roy Mukamel \\ 
	School of Psychological Science\\
	Tel Aviv University, Israel.
	\and Jelle Goeman \\ 
	Department of Medical Statistics and Bioinformatics, \\
	Leiden University Medical Center, The Netherlands.
	}

\newcommand{\set}[1]{\{ #1 \}} 
\newcommand{\indicator}[1]{\mathcal{I}{\set{#1}}} 
\newcommand{\reals}{\mathbb{R}} 
\newcommand{\acc}{\mathcal{E}} 
\newcommand{\accEstim}{\hat{\mathcal{E}}} 
 
\newcommand{\hyp}{\algo_{\data}} 
\newcommand{\hypFun}[2]{\algo_{#1}(#2)} 

\newcommand{\rv}[1]{\mathbf{#1}} 
\newcommand{\x}{\rv x} 
\newcommand{\y}{\rv y} 
\newcommand{\gauss}[1]{\mathcal{N}\left(#1\right)} 
\newcommand{\gaussp}[2]{\mathcal{N}_{#1}\left(#2\right)} 

\newcommand{\R}{\textsf{R }}
\newcommand{\algo}{\mathcal{A}}
\newcommand{\data}{\mathcal{S}}
\newcommand{\measure}{\mathcal{P}}
\newcommand{\measuren}{\measure_\data}
\newcommand{\union}{\cup}

\title{Better-Than-Chance Classification for Signal Detection}

\begin{document}

\maketitle

\begin{abstract}
The estimated accuracy of a classifier is a random quantity with variability. 
A common practice in supervised machine learning, is thus to test if the estimated accuracy is significantly better than chance level.
This method of signal detection is particularly popular in neuroimaging and genetics.
We provide evidence that using a classifier's accuracy as a test statistic can be an underpowered strategy for finding differences between populations, compared to a bona-fide statistical test.
It is also computationally more demanding than a statistical test. 
Via simulation, we compare test statistics that are based on classification accuracy, to others based on multivariate test statistics. 
We find that probability of detecting differences between two distributions is lower for accuracy based statistics.
We examine several candidate causes for the low power of accuracy tests. 
These causes include: the discrete nature of the accuracy test statistic, the type of signal accuracy tests are designed to detect, their inefficient use of the data, and their regularization. 
When the purposes of the analysis is not signal detection, but rather, the evaluation of a particular classifier, we suggest several improvements to increase power. 
In particular, to replace V-fold cross validation with the Leave-One-Out Bootstrap.
\end{abstract}


\section{Introduction}
\label{sec:introduction}

Many neuroscientists and geneticists detect signal by fitting a classifier and testing whether it's prediction accuracy is better than chance. 
The workflow consists of fitting a classifier, and estimating its predictive accuracy using cross validation. 
Given that the cross validated accuracy is a random quantity, it is then common to test if the cross validated accuracy is significantly better than chance using a permutation test.  
Examples in the neuroscientific literature include \citet{golland_permutation_2003,pereira_machine_2009,schreiber2013statistical,varoquaux_assessing_2016}, and especially the recently popularized \emph{multivariate pattern analysis} (MVPA) framework of \citet{kriegeskorte_information-based_2006}.
For examples in the genetics literature see for example
\citet{golub_molecular_1999,slonim_class_2000,radmacher_paradigm_2002,mukherjee_estimating_2003,juan_prediction_2004,jiang_calculating_2008}.

To fix ideas, we will adhere to a concrete example.
In \cite{gilron_quantifying_2016}, the authors seek to detect brain regions that encode differences between vocal and non-vocal stimuli. 
Following the MVPA workflow, the localization problem is cast as a supervised learning problem: if the type of the stimulus can be predicted from the brain's activation pattern significantly better than chance, then a region is declared to encode vocal/non-vocal information. 
We call this an \emph{accuracy test}, because it uses the prediction accuracy as a test statistic. 

This same signal detection task can also be approached as a multivariate test.
Inferring that a region encodes vocal/non-vocal information, is essentially inferring that the spatial distribution of brain activations is different given a vocal/non-vocal stimulus. 
As put in \cite{pereira_machine_2009}: 
\begin{quote}
... the problem of deciding whether the classifier learned to discriminate the classes can be subsumed into the more general question as to whether there is evidence that the underlying distributions of each class are equal or not.
\end{quote}
A practitioner may thus approach the signal detection problem with a two-group location test such as Hotelling's $T^2$ \citep{anderson_introduction_2003}.
Alternatively, if the size of the brain's region of interest is large compared to the number of observations, so that the spatial covariance cannot be fully estimated, then a high dimensional version of Hotelling's test can be called upon.
Examples of high dimensional multivariate tests include \cite{schafer_shrinkage_2005}, \cite{goeman2006testing}, or \cite{srivastava_multivariate_2007} .
For brevity, and in contrast to \emph{accuracy tests}, we will call these \emph{location tests}, because they test for the equality of location of two multivariate distributions.

At this point, it becomes unclear which is preferable: a location test or an accuracy test?
The former with a heritage dating back to \cite{hotelling_generalization_1931}, and the latter being extremely popular, as the $1,170$ citations\footnote{GoogleScholar. Accessed Aug 2017.} of \cite{kriegeskorte_information-based_2006} suggest. 

The comparison between location and accuracy tests was precisely the goal of \cite{ramdas_classification_2016}, who compared Hotelling's $T^2$ location test to \emph{Fisher's linear discriminant analysis} (LDA) accuracy test. 
By comparing the rates of convergence of the power of each statistic, \cite{ramdas_classification_2016} concluded that accuracy and location tests are rate equivalent. 
Rates, however, are only a first stage when comparing test statistics. 

Asymptotic relative efficiency measures (ARE) are typically used by statisticians to compare between rate-equivalent test statistics \citep{vaart_asymptotic_1998}.
ARE is the limiting ratio of the sample sizes required by two statistics to achieve similar power. 
\cite{ramdas_classification_2016} derive the asymptotic power functions of the two test statistics, which allows to compute the ARE between Hotelling's $T^2$ (location) test and Fisher's LDA (accuracy) test.
Theorem~14.7 of \cite{vaart_asymptotic_1998} relates asymptotic power functions to ARE.
Using this theorem and the results of \cite{ramdas_classification_2016} we deduce that the ARE is lower bounded by $2 \pi \approx 6.3$. 
This means that Fisher's LDA requires at least $6.3$ more samples to achieve the same (asymptotic) power as the $T^2$ test. 
In this light, the accuracy test is remarkably inefficient compared to the location test.  
For comparison, the t-test is only $1.04$ more (asymptotically) efficient than Wilcoxon's rank-sum test \citep{lehmann_parametric_2009}, so that an ARE of $6.3$ is strong evidence in favor of the location test. 

Before discarding accuracy tests as inefficient, we recall that \cite{ramdas_classification_2016} analyzed a \emph{half-sample} holdout. 
The authors conjectured that a leave-one-out approach, which makes more efficient use of the data, may have better performance. 
Also, the analysis in \cite{ramdas_classification_2016} is asymptotic. 
This eschews the discrete nature of the accuracy statistic, which we will show to have  crucial impact. 
Since typical sample sizes in neuroscience are not large, we seek to study which test is to be preferred in finite samples, and not only asymptotically.
Our conclusion will be quite simple: {\em location tests typically have more power than accuracy tests, are easier to implement, and interpret.}

Our statement rests upon the observation that with typical sample sizes, the accuracy test statistic is highly discrete. 
Permutation testing with discrete test statistics are known to be conservative \citep{hemerik_exact_2014}, since they are insensitive to mild perturbations of the data, and cannot exhaust the permissible false positive rate. 
As put by Prof. Frank Harrell in \textsf{CrossValidated\footnote{A Q\&A website for statistical questions: \url{http://stats.stackexchange.com/questions/17408/how-to-assess-statistical-significance-of-the-accuracy-of-a-classifier}}} post back in $2011$:
\begin{quote}
	... your use of proportion classified correctly as your accuracy score. This is a discontinuous improper scoring rule that can be easily manipulated because it is arbitrary and insensitive.
\end{quote}

The degree of discretization is governed by the number of samples. 
In our example from \citet{gilron_quantifying_2016}, the classification accuracy is computed using $40$ samples, so that the test statistic may assume only $40$ possible values. 
This number of samples is not unusual in an neuroimaging study. 

Power loss due to discretization is further aggravated if the test statistic is highly concentrated. 
For an intuition consider the usage of the \emph{resubstitution accuracy}, a.k.a.\ the \emph{train error}, or \emph{empirical risk}, as a test statistic. 
Resubstitution accuracy is the accuracy of the classifier evaluated on the training set.
If data is high dimensional, the resubstitution accuracy will be very high due to over fitting. 
In a very high dimensional regime, the resubstitution accuracy may be as high as $1$ for the observed data \cite[Theorem 1]{mclachlan_bias_1976}, but also for any permutation.
The concentration of resubstitution accuracy near $1$, and its discretization, render this test completely useless, with power tending to $0$ for any (fixed) effect size, as the dimension of the model grows.

To compare the power of accuracy tests and location tests in finite samples, we study a battery of test statistics by means of simulation. 
We start with formalizing the problem in Section~\ref{sec:problem_setup}.
The main findings are reported in Sections~\ref{sec:results}, and \ref{sec:example}.
A discussion follows.

\section{Problem setup}
\label{sec:problem_setup}

\subsection{Multivariate Testing}

Let $y \in \mathcal{Y}$ be a class encoding. 
Let $x \in \mathcal{X}$ be a $p$ dimensional feature vector. 
In our vocal/non-vocal example we have $\mathcal{Y}=\set{0,1}$ and $p=27$, the number of voxels in a brain region so that $\mathcal{X}=\reals^{27}$. 

Denoting a dataset by $\data:=\{(x_i,y_i)\}_{i=1}^n$, a multivariate test amounts to testing whether the distribution of $x$ given $y=1$ is the same as $x$ given $y=0$. 
For example, we can test whether multivariate voxel activation patterns ($x$) are similarly distributed when given a vocal stimulus ($y=1$) or a non-vocal one ($y=0$).
The tests are calibrated to have a fixed false positive rate ($\alpha=0.05$).
The comparison metric between statistics is power, i.e., the probability to infer that $x|y=1$ is not distributed like $x|y=0$.

\subsection{From a Test Statistic to a Permutation Test}

The multivariate tests we will be considering rely on fixing some test statistic, and comparing it to it's permutation distribution. 
The tests differ in the statistic they employ.
Our comparison metric is their power, i.e., their true positive rate. 
We adhere to permutation tests and not parametric inference because our problems of interest are typically high-dimensional. 
This means that $n \gg p$ does not hold, and central limit laws do not apply.
Because we focus on two-group testing under an independent sampling assumption, we know that a label-switching permutation test is valid even if possibly conservative. 
The sketch of our permutation test is the following: \newline
(a) Fix a test statistic $\mathcal{T}$ with a right tailed rejection region. \newline
(b) Sample a random permutation of the class labels, $\pi(y)$. \newline
(c) Permute labels and recompute the statistic $\mathcal{T}_\pi$. \newline
(d) Repeat (a)-(c) $R$ times. \newline
(e) The permutation p-value is the proportion of  $\mathcal{T}_\pi$ larger than the observed $\mathcal{T}$. Formally: 
$\mathbb{P}\{\mathcal{T}_\pi \geq \mathcal{T}\}:=\frac{1}{R} \sum_{\pi} I\{\mathcal{T}_\pi \geq \mathcal{T}\}$.\newline
(f) Declare classes differ if the permutation p-value is smaller than $\alpha$, which we set to $\alpha=0.05$.
\bigskip

We now detail the various test statistics that will be compared.

\subsection{Location Tests and Hotelling's $T^2$}
The most prevalent interpretation of ``$x|y=1$ is not distributed like $x|y=0$'' is to assume they differ in means. 
In his seminal work, \citet{hotelling_generalization_1931} has proposed the $T^2$ test statistic for testing the equality in means of two multivariate distributions. 
Using our notations this statistic is proportional to the difference between group means, measured with the Mahalanobis norm: 
\begin{align}
	T^2 \propto \; (\bar{x}_{y=1}-\bar{x}_{y=0})'\; \hat{\Sigma}^{-1} \;(\bar{x}_{y=1}-\bar{x}_{y=0}), 
\end{align}
where $\bar{x}_{y=j}$ is the $p$-vector of means in the $y=j$ group, and $\hat{\Sigma}$ is a pooled covariance estimator.
Perhaps more intuitively, $T^2$ is Euclidean norm of the mean difference vector, but after transforming to decorrelated scales. 
For more background see, for example, \cite{anderson_introduction_2003}.

The major difficulty with these multivariate tests is that $\Sigma$ has $p(p+1)/2$ free parameters, so that $n$ has to be very large to apply these tests.
If $n$ is not much larger than $p$, or in low signal-to-noise (SNR), the test is very low powered, as shown by \cite{bai1996effect}. 
In these cases, high dimensional versions of the $T^2$ should be applied, which essentially regularize the estimator of $\Sigma$, thus reducing the dimensionality of the problem and improving the SNR and power.

\subsection{Prediction Accuracy as a Test Statistic}
An accuracy test amounts to using a predictor's accuracy as a test statistic.  

A predictor\footnote{Known as a \emph{hypothesis} in the machine learning literature.}, $\hyp:\mathcal{X} \to \mathcal{Y}$, is the output of a learning algorithm $\algo$ when applied to the dataset $\data$. 
The accuracy of predictor\footnote{Known as (the complement of) the \emph{test error} in \cite{hastie_elements_2003}}, $\acc_{\hyp}$, is defined as the probability of $\hyp$ making a correct prediction. 
The accuracy of an algorithm\footnote{Known as (the complement of) the \emph{expected test error} in \cite{hastie_elements_2003}}, $\acc_{\algo}$, is defined as the expected accuracy over all possible data sets $\data$. 
Formalizing, we denote by $\measure$ the probability measure of $(x, y)$, and by $\measuren$ the joint probability measure of the sample $\data$. 
We can then write 
\begin{align}
	\acc_{\hyp}:=\int_{(x,y)} \indicator{\hyp(x)=y} \; d\measure,
\end{align}
and
\begin{align}
	\acc_{\algo}:=\int_\data \acc_{\algo_\data} \; d\measuren,
\end{align}
where $\indicator{A}$ is the indicator function\footnote{Mutatis mutandis for continous $y$.} of the set $A$. 

Denoting an estimate of $\acc_{\hyp}$ by $\accEstim_{\hyp}$, and $\acc_{\algo}$ by $\accEstim_{\algo}$, a statistically significant ``better than chance'' estimate of either, is evidence that the classes are distinct. 

Two popular estimates of $\accEstim_{\algo}$ are the \emph{resubstitution estimate}, and the V-fold Cross Validation (CV) estimate.
\begin{definition}[Resubstitution estimate]
\label{def:resubstitution}
The resubstitution accuracy estimator of a learning algorithm $\algo$, denoted $\accEstim_{\algo}^{Resub}$,  is defined as
\begin{align}
	\accEstim_{\algo}^{Resub} := \frac 1n \sum_{i=1}^{n} \indicator{\hypFun{\data}{x_i}=y_i}.
\end{align}
\end{definition}

\begin{definition}[V-fold CV estimate]
\label{def:v-fold}
Denoting by $\data^{v}$ the $v$'th partition, or \emph{fold}, of the dataset, and by $\data^{(v)}$ its complement, so that $\data^{v} \union \data^{(v)}=\union_{v=1}^V \data^{v}=\data$, the V-fold CV accuracy estimator, denoted $\accEstim_{\algo}^{Vfold}$, is defined as 	
\begin{align}
	\accEstim_{\algo}^{Vfold} := 
	\frac 1V \sum_{v=1}^{V} \frac{1}{|\data^v|} \sum_{i \in \data^{v}} \indicator{\hypFun{\data^{(v)}}{x_i}=y_i},
\end{align}
where $|A|$ denotes the cardinality of a set $A$.
\end{definition}

\subsection{How to Estimate Accuracies?}
\label{sec:considerations}

Estimating $\accEstim_{\algo}$ requires the following design choices: 
Should it be cross-validated and how? 
If cross validating using V-fold CV then how many folds? 
Should the folding be balanced?
If estimation is part of a permutation test: should the data be refolded after each permutation? 

We will now address these questions while bearing in mind that unlike the typical supervised learning setup, we are not interested in an unbiased estimate of $\acc_{\algo}$, but rather in the detection of its departure from chance level. 

\paragraph{Cross validate or not?}
For the purpose of statistical testing, bias in $\hat\acc_{\algo}$ is not a problem, as long as it does not invalidate the error rate guarantees. 
The underlying intuition is that if the same bias is introduced in all permutations, it will not affect the properties of the permutation test. 
We will thus be considering both cross validated accuracies, and resubstitution accuracies.

\paragraph{Balanced folding?}
The standard practice in V-fold CV is to constrain the data folds to be balanced, i.e. stratified \citep[e.g.][]{ojala_permutation_2010}.
This means that each fold has the same number of examples from each class. 
We will report results with both balanced and unbalanced data foldings.

\paragraph{Refolding?}
In V-fold CV, \emph{folding} the data means assigning each observation to one of the $V$ data folds. 
The standard practice in neuroimaging is to permute labels and refold the data after each permutation. 
This is done because permuting labels will unbalance the original balanced folding.
We will adhere to this practice due to its popularity, even though it is computationally more efficient to permute features\footnote{The difference between permuting labels or features is in the mapping to folds. When permuting features, the \textit{label} assignment to folds is fixed. When permuting labels, the \textit{feature} assignment to folds is fixed.} instead of labels, as done by \citet{golland_permutation_2005}.

\paragraph{How many folds?}
Different authors suggest different rules for the number of folds. 
We fix the number of folds to $V=4$, and do dot discuss the effect of $V$ because we will ultimately show that V-fold CV is dominated by other cross-validation procedures, and thus, never recommended. 

\bigskip

Table~\ref{tab:collected} collects an initial battery of tests we will be comparing. 
\begin{tcolorbox}
\centering
\begin{tabular}{l|c|c|c}
Name & Algorithm & Resampling & Parameters\\ 
\hline
\hline
Oracle & Hotelling & Resubstitution & -- \\ 
Hotelling & Hotelling & Resubstitution & -- \\ 
Hotelling.shrink & Hotelling & Resubstitution & -- \\ 
Goeman & Hotelling & Resubstitution & -- \\ 
sd & Hotelling & Resubstitution & -- \\ 
lda.CV.1 	& LDA & V-fold 			&  -- \\ 
lda.noCV.1 	& LDA & Resubstitution 	&  --\\ 
svm.CV.1 	& SVM & V-fold 		    & cost=$10$ \\ 
svm.CV.2 	& SVM & V-fold 		    & cost=$0.1$ \\ 
svm.noCV.1 	& SVM & Resubstitution  & cost=$10$ \\ 
svm.noCV.2 	& SVM & Resubstitution  & cost=$0.1$ \\ 
\end{tabular} 
\captionsetup{type=table}
\caption{\footnotesize
This table collects the various test statistics we will be studying. 
Location tests include: \textit{Oracle}, \textit{Hotelling}, \textit{Hotelling.shrink}, \textit{Goeman}, and \textit{sd}.
\textit{Oracle} is the same as Hotelling's $T^2$, only using the generative covariance, and not an estimated one.
\textit{Hotelling} is the classical two-group $T^2$ statistic \citep{anderson_introduction_2003}. 
\textit{Hotelling.shrink} is a high dimensional version of $T^2$, with the regularized covariance from \citet{schafer_shrinkage_2005}. 
\textit{Goeman} and \textit{sd} are other high dimensional versions of the $T^2$, from \cite{goeman2006testing} and \cite{srivastava_testing_2013}.
The rest of the tests are accuracy tests, with details given in the table. 
For example, \textit{svm.CV.2} is a linear SVM, with V-fold cross validated accuracy, and cost parameter set at $0.1$ \citep{meyer_e1071:_2015}.
Another example is \textit{lda.noCV.1}, which is Fisher's LDA, with a resubstituted accuracy estimate.}
\label{tab:collected}
\end{tcolorbox}

\section{Results}
\label{sec:results}
We now compare the power of our various statistics in various configurations. 
We do so via simulation.
The basic simulation setup is presented in Section~\ref{sec:simulation_details}.
Following sections present variations on the basic setup.
The \R code for the simulations can be found in \url{http://www.john-ros.com/permuting_accuracy/}.

\subsection{Basic Simulation Setup}
\label{sec:simulation_details}

Each simulation is based on $1,000$ replications. 
In each replication, we generate $n$ i.i.d. samples from a shift class 
\begin{align}
\label{eq:distribution}
	\x_i = \mu \y_i + \eta_i,
\end{align}
where $\y_i \in \mathcal{Y}=\set{0,1}$ encodes the class of subject $i$, $\mu$ is a $p$-dimensional shift vector, the noise $\eta_i$ is distributed as $\gaussp{p}{0,\Sigma}$, the sample size $n=40$, and the dimension of the data is $p=23$. 
The covariance $\Sigma=I$. 
In this basic setup, reported in Figure~\ref{fig:simulation_1}, the shift effect is captured by $\mu$. 
Shifts are equal in all $p$ coordinates of $\mu$.
With $e$ being a $p$-vector of ones, then $\mu:=c \, e$. 
We will use $c$ to index the signal's strength, and vary it over $c \in \set{0,1/4,1/2}$.
The (squared) Euclidean and Mahalanobis norms of the signal are $\Vert \mu \Vert_2^2=\Vert \mu \Vert_\Sigma^2=c^2 p\approx \{0,1.4,5.7\}$.
These can be thought as the effect's size.

Having generated the data, we compute each of the test statistics in Table~\ref{tab:collected}.
For test statistics that require data folding, we used $4$ folds. 
We then compute a permutation p-value by permuting the class labels, and recomputing each test statistic. 
We perform $300$ such permutations. 
We then reject the ``$x|y=0$ distributed like $x|y=1$'' null hypothesis if the permutation p-value is smaller than $0.05$.
The reported power is the proportion of replication where the permutation p-value fell below $0.05$.

\subsection{False Positive Rate}
\label{sec:type_i}

We start with a sanity check. 
Theory suggests that all test statistics should control their false positive rate. 
Our simulations confirm this.
In all our results, such as Figure~\ref{fig:simulation_1}, we encode the null case, where no signal is present and $x|y=1$ has the same distribution as $x|y=0$, by a red circle. 
Since the red circles are always below the desired $0.05$ error rate then the false positive rate of all test statistics, in all simulations is controlled. 
We may thus proceed and compare the power of each test statistic.

\subsection{Power}
\label{sec:power}

Having established that all of the tests in our battery control the false positive rate, it remains to be seen if they have similar power-- especially when comparing location tests to accuracy tests. 

\begin{figure}[h]
	\centering
	\caption{
		The power of the permutation test with various test statistics. 
		The power on the $x$ axis. 
		Effects are color and shape coded. 
		The various statistics on the $y$ axis. 
		Their details are given in Table~\ref{tab:collected}. 
		Effects vary over $c=0$ (red circle), $c=1/4$ (green triangle), and $c=1/2$ (blue square). 
		Simulation details in Section~\ref{sec:simulation_details}.
		Cross-validation was performed with balanced and unbalanced data folding; see sub-captions.}	
	\label{fig:simulation_1}
	\begin{subfigure}{.5\textwidth}
		\centering
		\includegraphics[width=1\linewidth]{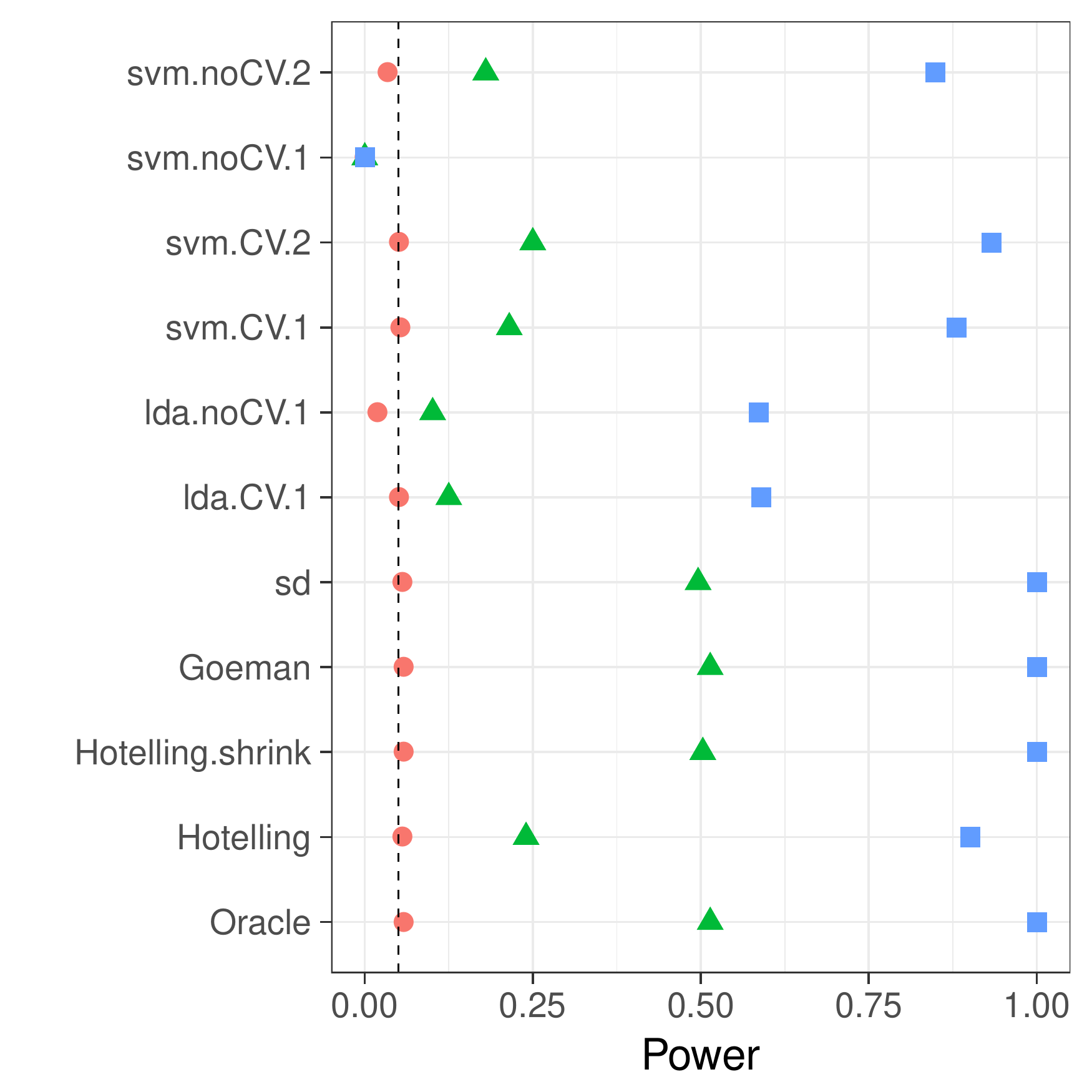}
		\caption{\textbf{Unbalanced V-fold CV.}} 
		\label{fig:file3}
	\end{subfigure}%
	\begin{subfigure}{.5\textwidth}
		\centering
		\includegraphics[width=1\linewidth]{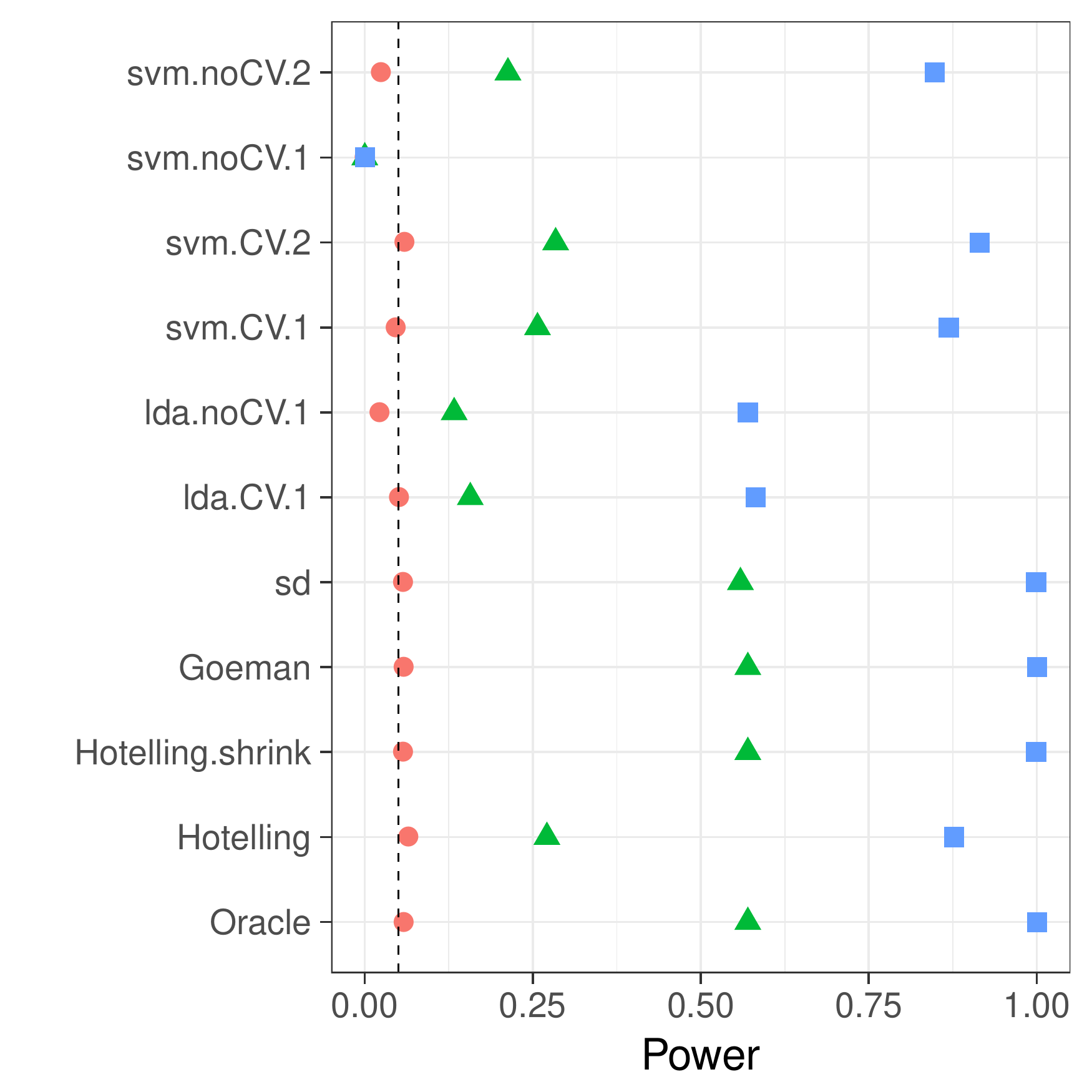}
		\caption{\textbf{Balanced V-fold CV.}} 
		\label{fig:file2}
	\end{subfigure}
\end{figure}

From Figure~\ref{fig:simulation_1} we learn that location tests are more powerful than accuracy tests.
This is particularly visible for intermediate signal strength (green triangle), and location tests \emph{Goeman}, \emph{sd} and \emph{Hotelling.shrink} defined in Table~\ref{tab:collected}.

\subsection{Tie Breaking}
\label{sec:ties}

As already stated in the introduction, the accuracy statistic is highly discrete. 
Especially the resubstitution accuracy tests. 
Discrete test statistics lose power by not exhausting the permissible false positive rate. 
A common remedy is a \emph{randomized test}, in which the rejection of the null is decided at random in a manner that exhausts the false positive rate. 
Formally, denoting by $\mathcal{T}$ the observed test statistic, by $\mathcal{T}_\pi$, its value after under permutation $\pi$, and by $\mathbb{P}\{A\}$ the proportion of permutations satisfying $A$ then the randomized version of our tests imply that if the permutation p-value, 
$\mathbb{P}\{\mathcal{T}_\pi \geq \mathcal{T}\}$, 
is greater than  $\alpha$ then we reject the null with probability 
$$ max\left\{\frac{\alpha - \mathbb{P}\{\mathcal{T}_\pi > \mathcal{T}\}}{\mathbb{P}\{\mathcal{T}_\pi = \mathcal{T}\}},0 \right\}.$$

\begin{figure}[ht]
	\centering
	\includegraphics[width=0.5\linewidth]{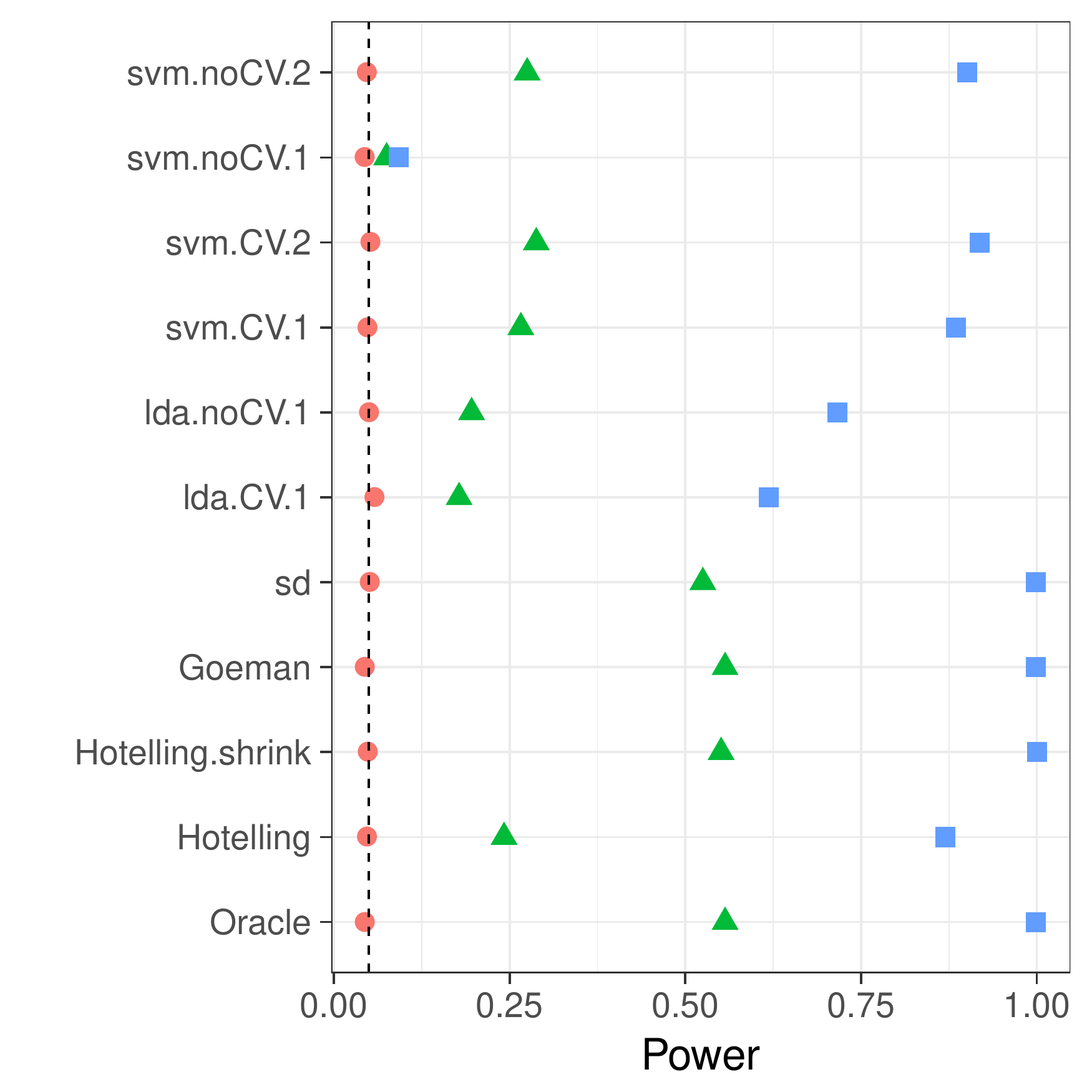}
	\caption{The same as Figure~\ref{fig:file2}, with random tie breaking.}
	\label{fig:file33}
\end{figure}

Figure~\ref{fig:file33} reports the same analysis as in Figure~\ref{fig:file2}, after allowing for random tie breaking. 
It demonstrates that the power disadvantage of accuracy tests, cannot be remedied by random tie breaking.

\subsection{Departure From Gaussianity}
The Neyman-Pearson Lemma (NPL) type reasoning that favors the location test over accuracy tests may fail when the data is not multivariate Gaussian, and Hotelling's $T^2$ statistic no longer a generalized-likelihood-ratio test. 

\begin{figure}[th]
		\centering
		\includegraphics[width=0.5\linewidth]{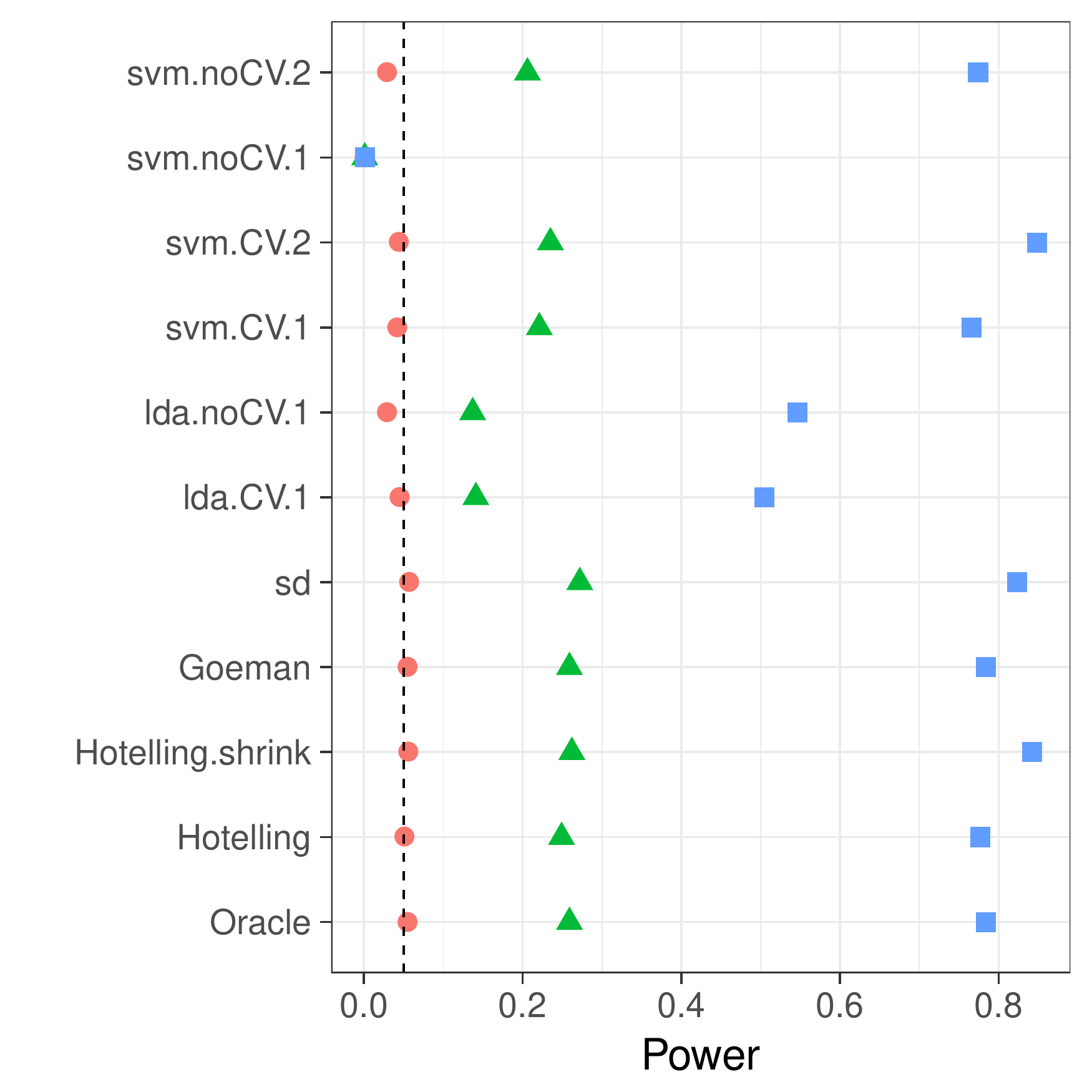}
		\caption{\textbf{Heavytailed.} $\eta_i$ is $p$-variate t, with $df=3$ .  } 
		\label{fig:t_null}
\end{figure}

To check this, we replaced the multivariate Gaussian distribution of $\eta$ in Eq.(\ref{eq:distribution}) with a heavy-tailed multivariate-$t$ distribution. 
In this heavytailed setup, the dominance of the location tests was preserved, even if less evident than in the Gaussian case (Figure~\ref{fig:t_null}).

\subsection{Departure from Sphericity}
\label{sec:dependence}

We now test the robustness of our results to the correlations in $x$. 
In terms of Eq.(\ref{eq:distribution}), $\Sigma$ will no longer be the identity matrix. 
Intuitively- both location tests and accuracy tests include the estimation of $\Sigma$, so that correlations should be accounted for. 
To keep the comparisons ``fair'' as the correlations vary, we kept $\Vert \mu \Vert_\Sigma:=\sqrt{\mu'\Sigma^{-1}\mu}$ fixed. 

Which test has more power: accuracy or location?
We address this question using various correlation structures.
We also vary the direction of the signal, $\mu$, and distinguish between signal in high variance principal component (PC) of $\Sigma$, and in the low variance PC. 

\begin{figure}[h]
	\centering
	\caption{Short memory, AR(1) correlation. 
	$\Sigma_{k,l}=\rho^{|k-l|}; \rho=0.6$}	
	\label{fig:dependence_1}
	\begin{subfigure}[t]{.4\textwidth}
	\centering
	\includegraphics[width=1\linewidth]{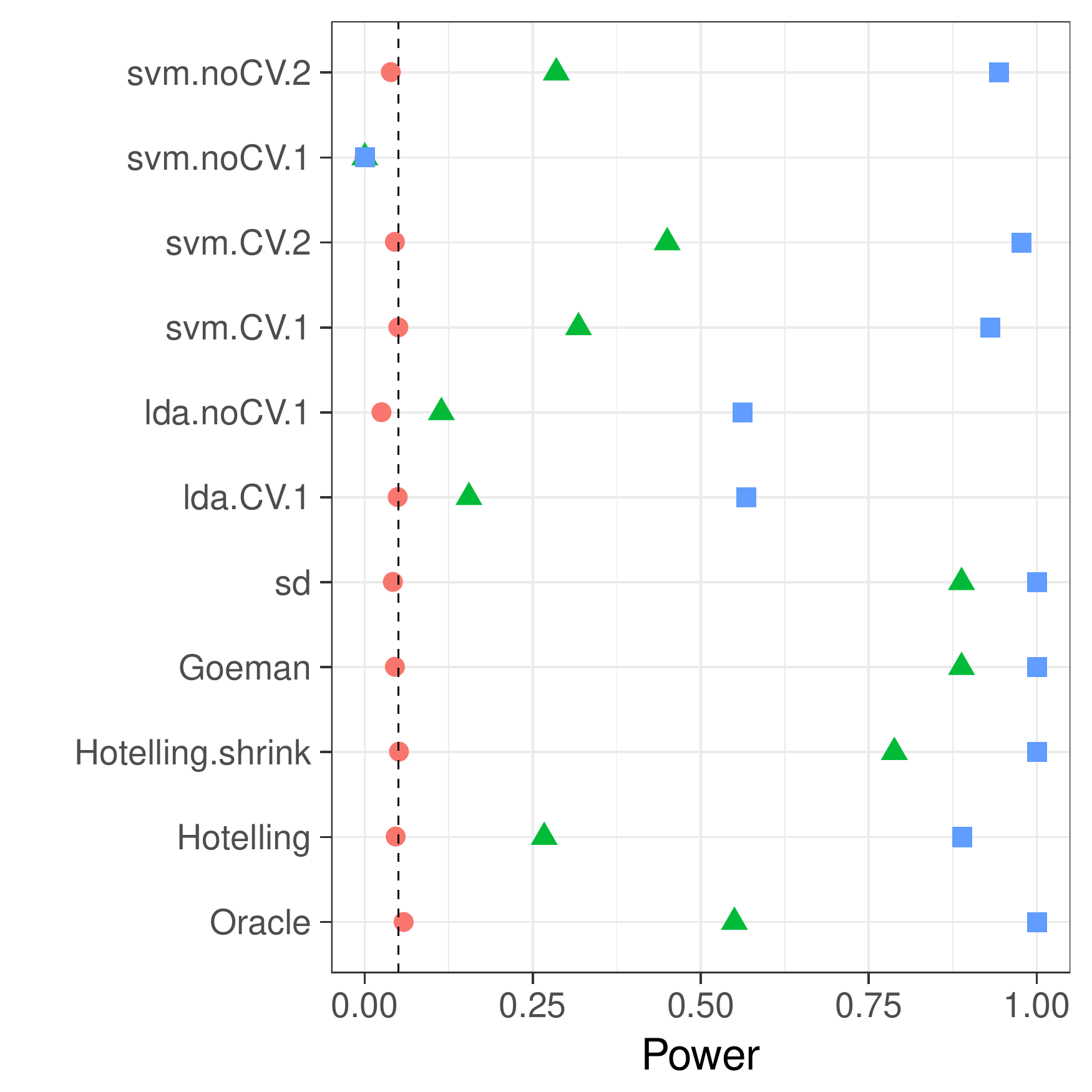}
	\caption{Signal in direction of highest variance PC of $\Sigma$.} 
	\label{fig:dependence_11}
	\end{subfigure}
	\begin{subfigure}[t]{.4\textwidth}
		\centering
		\includegraphics[width=1\linewidth]{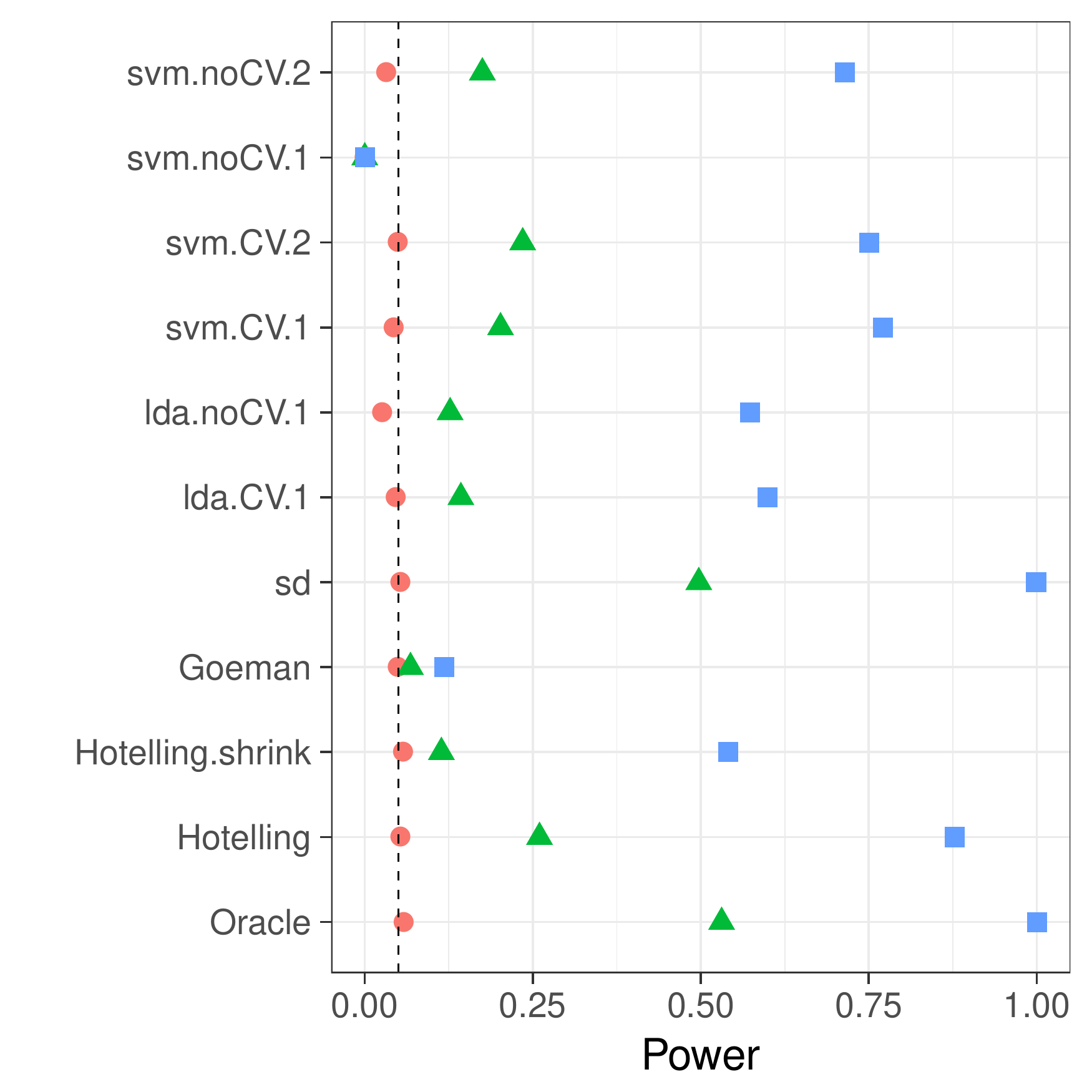}
		\caption{Signal in direction of lowest variance PC of $\Sigma$.} 
		\label{fig:dependence_12}
	\end{subfigure}
\end{figure}

\begin{figure}[h]
	\centering
	\caption{Long-memory Brownian motion correlation: $\Sigma=D^{-1} R D^{-1}$ where $D$ is diagonal with $D_{jj}=\sqrt{R_{jj}}$, and $R_{k,l}=\min\{k,l\}$.}	
	\label{fig:dependence_2}
	\begin{subfigure}[t]{.4\textwidth}
		\centering
		\includegraphics[width=1\linewidth]{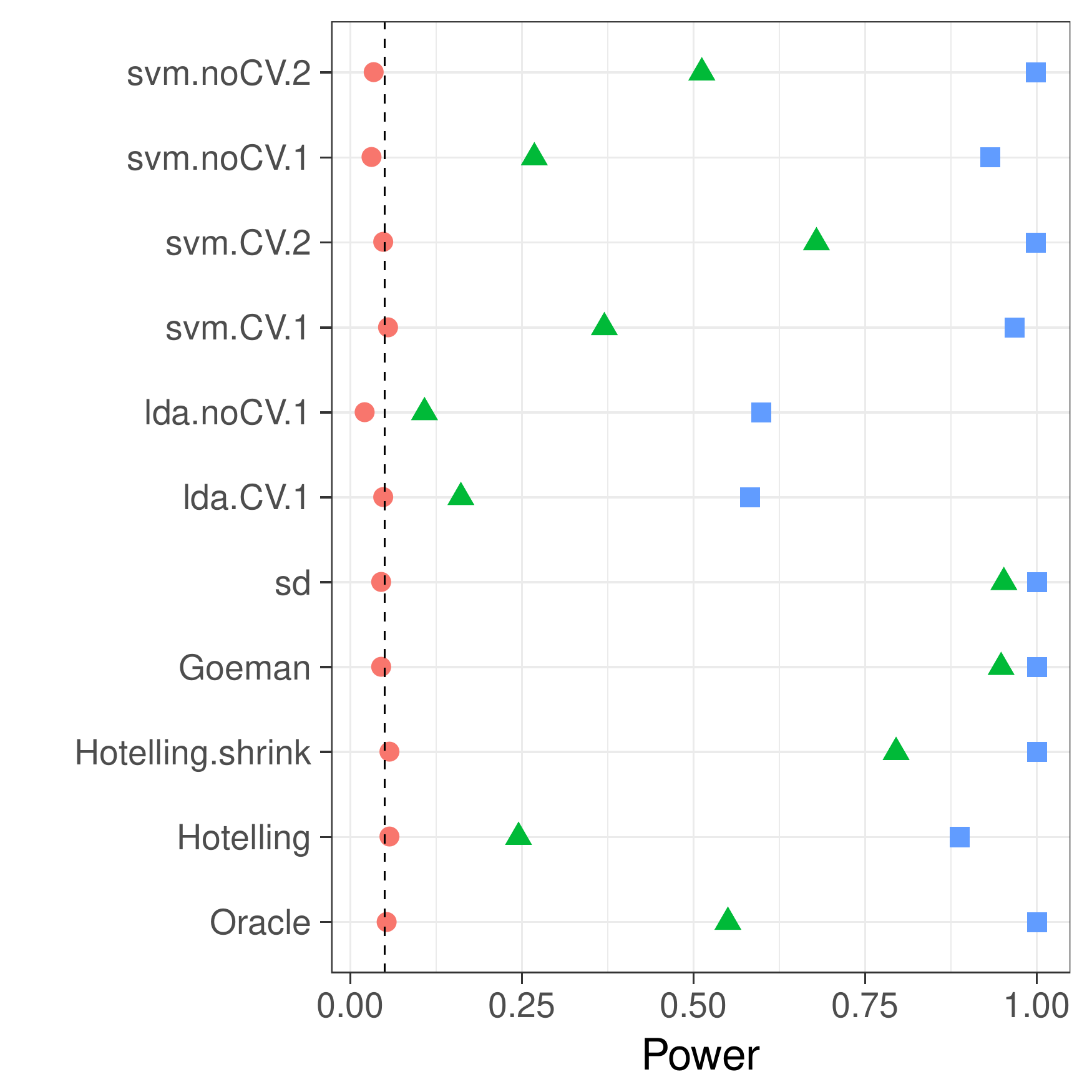}
		\caption{Signal in direction of highest variance PC of $\Sigma$.} 
		\label{fig:dependence_21}
	\end{subfigure}
	\begin{subfigure}[t]{.4\textwidth}
		\centering
		\includegraphics[width=1\linewidth]{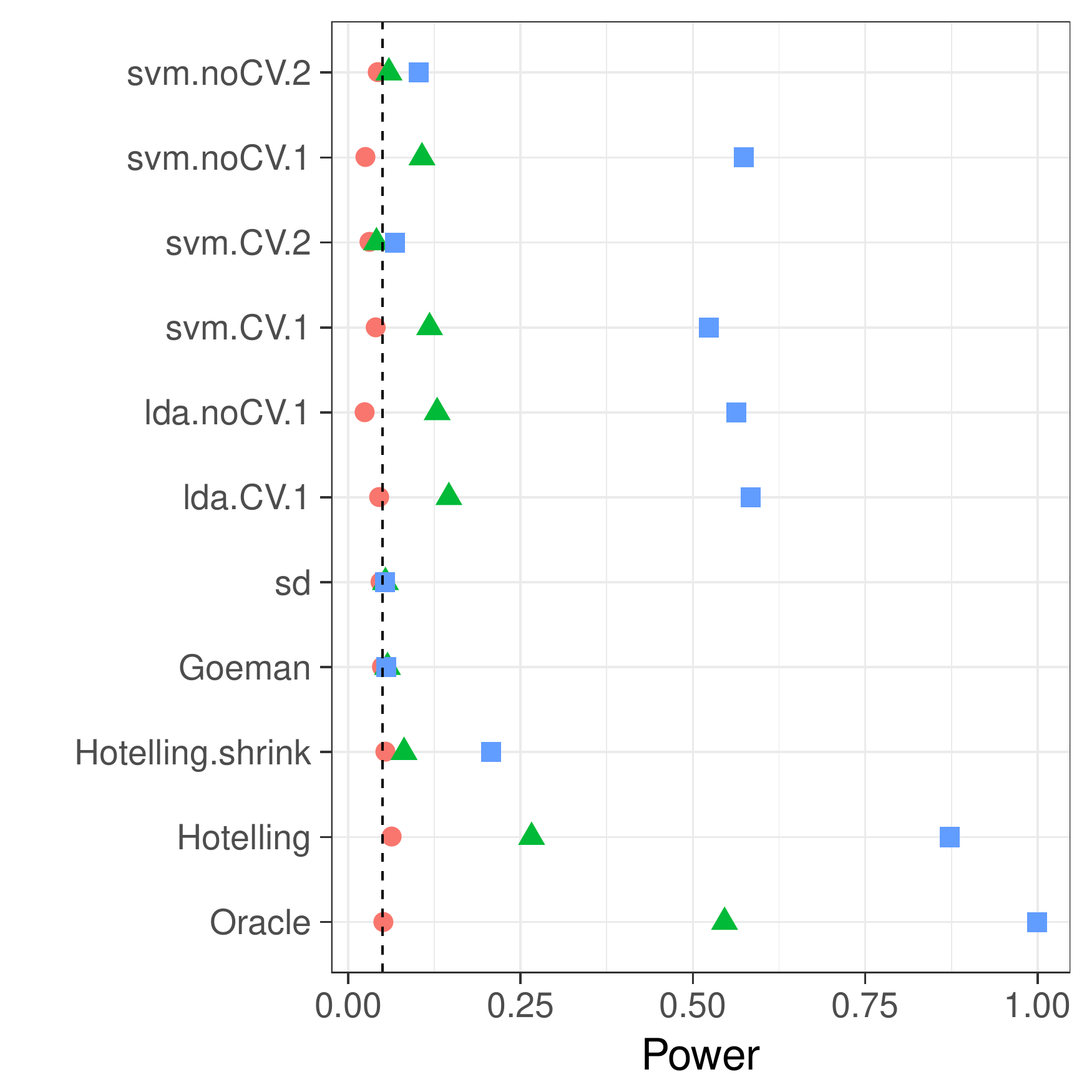}
		\caption{Signal in direction of lowest lowest variance PC of $\Sigma$.} 
		\label{fig:dependence_22}
	\end{subfigure}
\end{figure}

\begin{figure}[h]
	\centering
	\caption{Arbitrary Correlation. 
		$\Sigma=D^{-1} R D^{-1}$ where $D$ is diagonal with $D_{jj}=\sqrt{R_{jj}}$, and $R=A'A$ where $A$ is a Gaussian $p\times p$ random matrix with independent $\mathcal{N}(0,1)$ entries.
	}
	\label{fig:dependence_3}
	\begin{subfigure}[t]{.4\textwidth}
		\centering
		\includegraphics[width=1\linewidth]{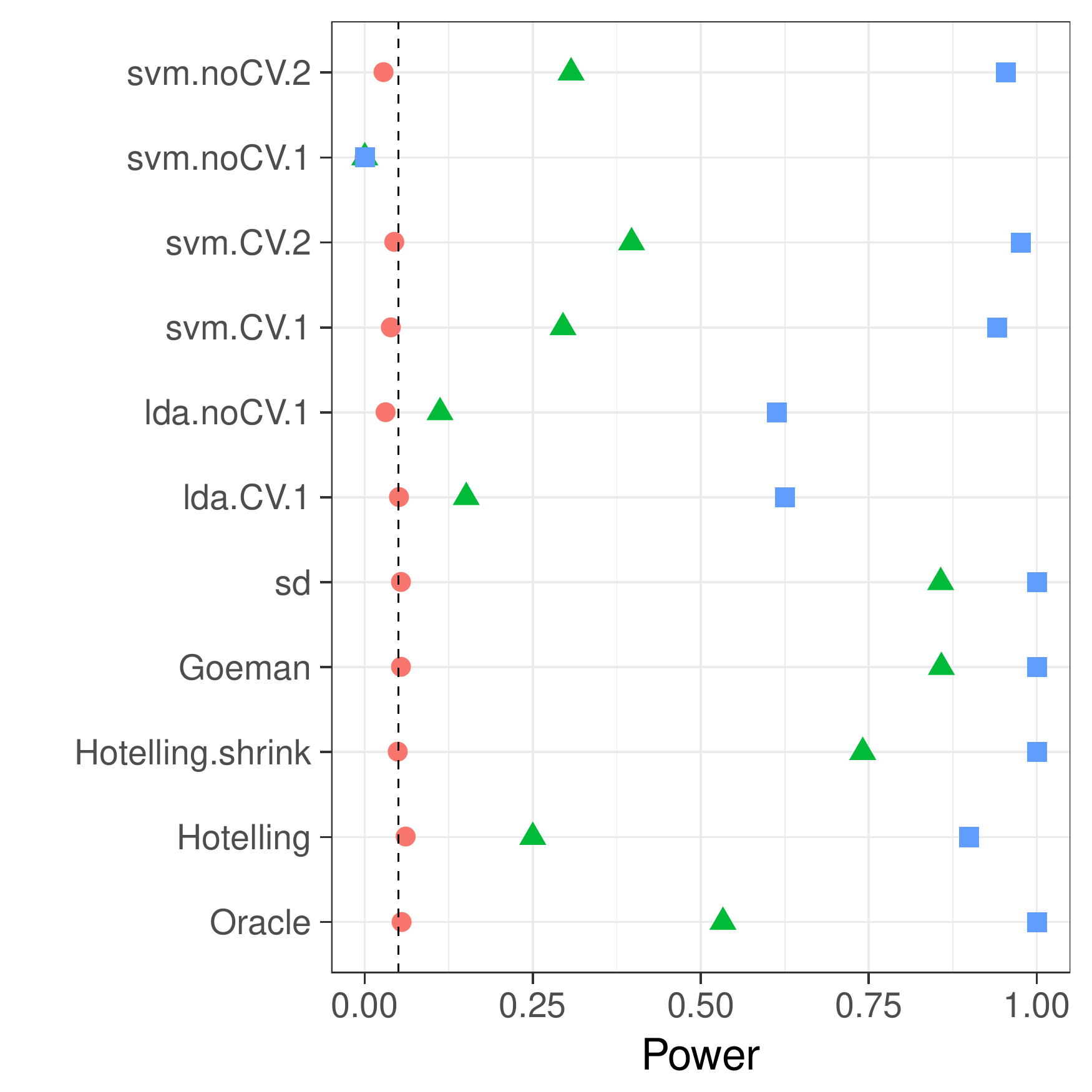}
		\caption{Signal in direction of highest variance PC of $\Sigma$.} 
		\label{fig:dependence_31}
	\end{subfigure}
	\begin{subfigure}[t]{.4\textwidth}
		\centering
		\includegraphics[width=1\linewidth]{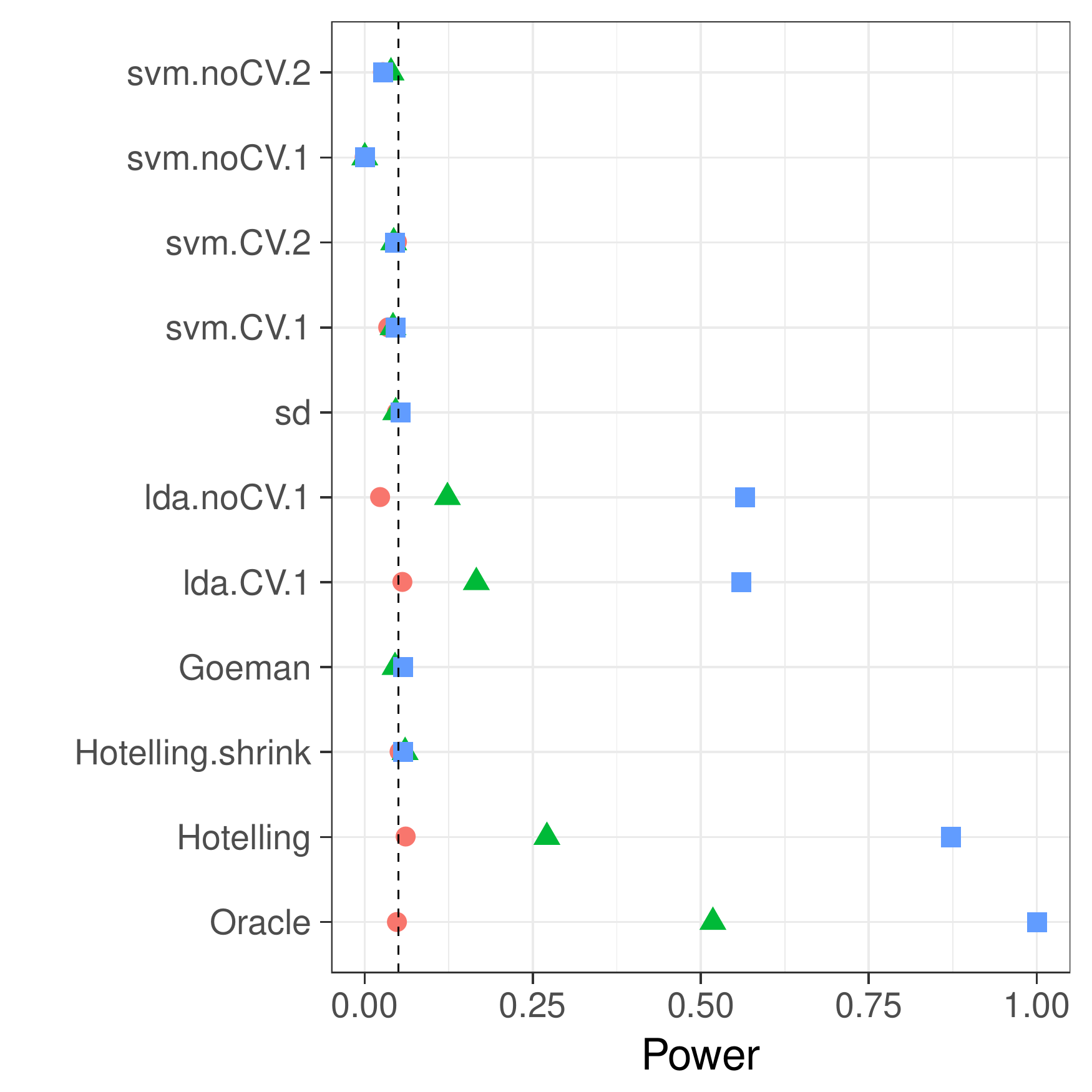}
		\caption{Signal in direction of lowest variance PC of $\Sigma$.} 
		\label{fig:dependence_32}
	\end{subfigure}
\end{figure}

The simulation results reveal some non trivial phenomena.
First, when the signal is in the direction of the high variance PC, the high dimensional location tests are far superior than accuracy tests. 
This holds true for various correlation structures: the short memory correlations of $AR(1)$ in Figure~\ref{fig:dependence_11}, the long memory correlations of a Brownian motion in Figure~\ref{fig:dependence_21}, and the arbitrary correlation in Figure~\ref{fig:dependence_31}.

When the signal is in the direction of the low variance PC, a different phenomenon appears.
There is no clear preference between location or accuracy tests.
Instead the non-regularized tests are the clear victors. 
This holds true for various correlation structures: the short memory correlations of $AR(1)$ in Figure~\ref{fig:dependence_12}, the long memory correlations of a Brownian motion in Figure~\ref{fig:dependence_22}, and the arbitrary correlation in Figure~\ref{fig:dependence_32}.
We attribute this phenomenon to the bias introduced by the regularization, which masks the signal.
This matter is further discussed in Section~\ref{sec:fix_snr}.

\subsection{Departure from Homoskedasticity}

Our previous simulations assume variables a have unit variance. 
The heteroskedastic case, where difference coordinates have different variance, is of lesser importance, since we can typically normalize the variable-wise variance. 
Some test statistics have built-in variance normalization, and are known as \emph{scalar invariant}. 
The \emph{sd} test statistic is scalar invariant. 
Statistics that are not scalar-invariant such as the \emph{Goeman} statistic, will give less importance to high-variance directions than to low-variance directions. 

\begin{figure}[h]
	\centering
	\caption{Heteroskedasticity: $\Sigma$ is diagonal with $\Sigma_{jj}=j$.}	
	\label{fig:heteroskedastic}	
	\begin{subfigure}[t]{.4\textwidth}
		\centering
		\includegraphics[width=1\linewidth]{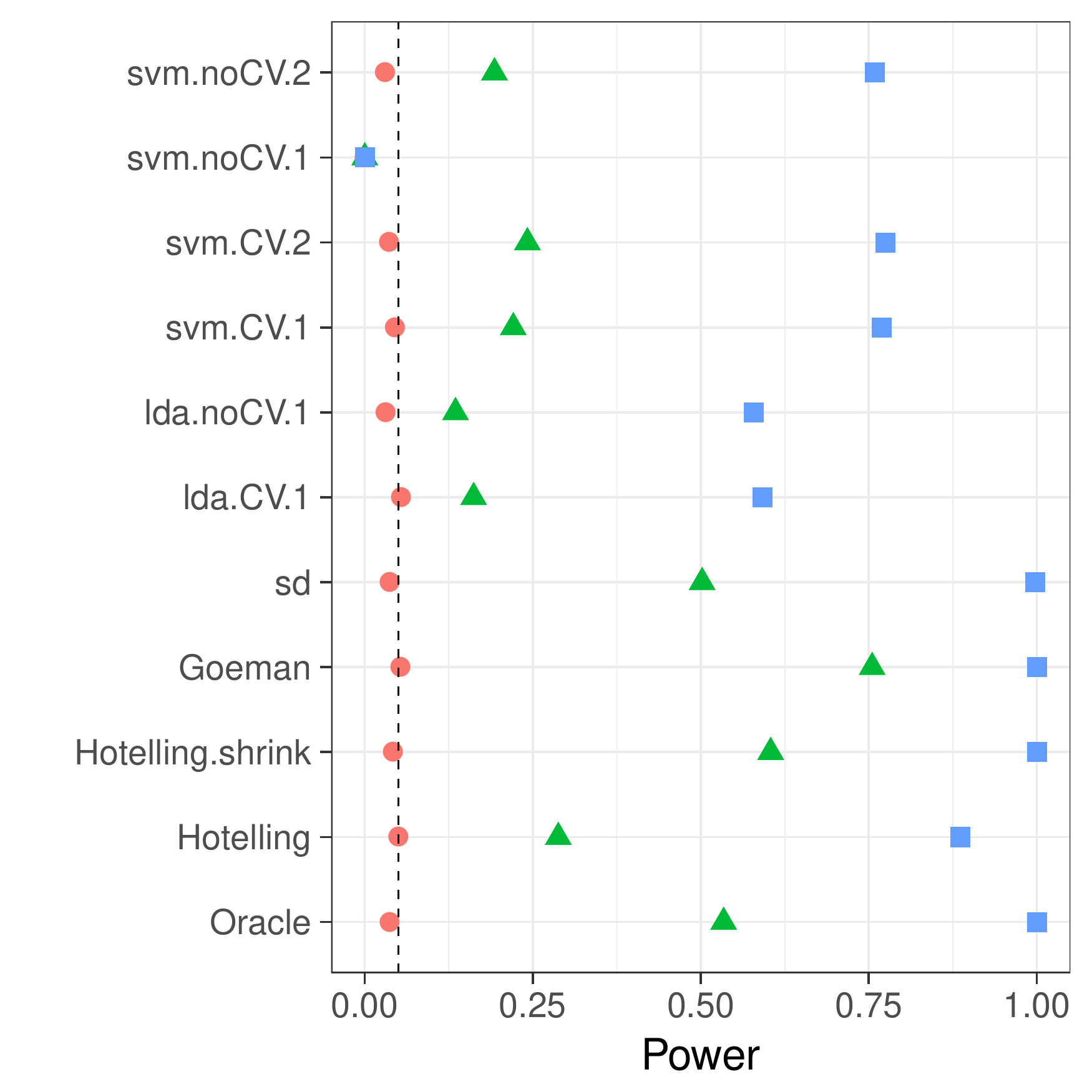}
		\caption{$\mu$ in the high variance PC of $\Sigma$.}  
		\label{fig:heteroskedastic_11}	
	\end{subfigure}
	\begin{subfigure}[t]{0.4\textwidth}
		\centering
		\includegraphics[width=1\linewidth]{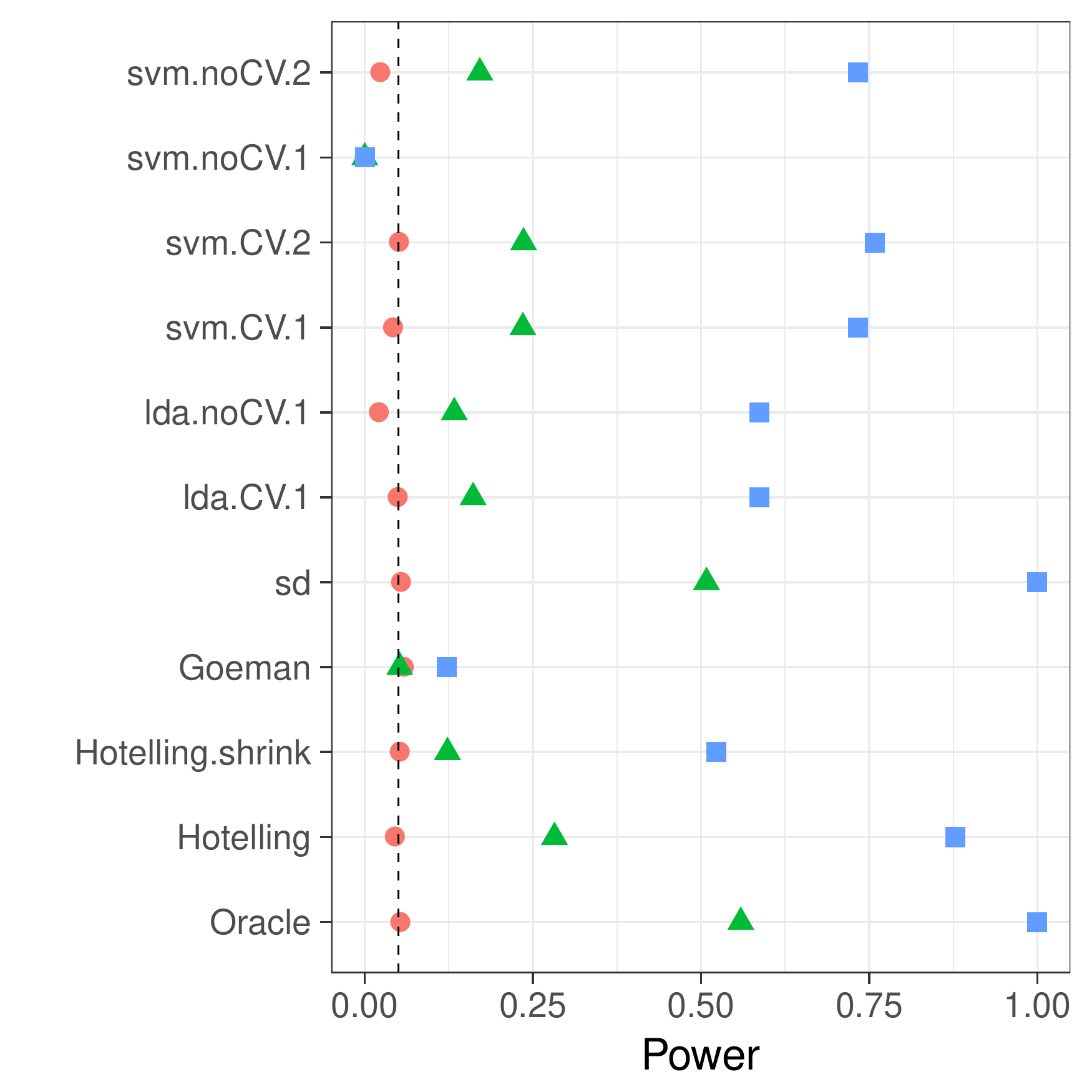}
		\caption{$\mu$ in the low variance PC of $\Sigma$.}  
		\label{fig:heteroskedastic_12}	
	\end{subfigure}
\end{figure}

In Figure~\ref{fig:heteroskedastic_11} we see that as before, location tests dominate accuracy tests.
For the first time, we can see the difference between the scalar-invariant \emph{sd} and \emph{Goeman}: the latter gaining power by focusing on low variance coordinates. Since the signal's magnitude is the same in all coordinates, \emph{Goeman} gains power by putting emphasis where it is needed.

When the signal is in the low variance PC, \emph{Goeman} puts emphasis on variables which carry little signal.
For this reason it has less power than \emph{sd}, as seen in Figure~\ref{fig:heteroskedastic_12}.

\subsection{Departure from V-fold CV}
\label{sec:bootstrap}

Intuition suggests we may alleviate the discretization of the accuracy test statistic by replacing the V-fold CV, and resampling \emph{with replacement}.
The discretization of the accuracy statistic is governed by the number of samples in the union of test sets.
For V-fold CV, for instance, the accuracy may assume as many values as the sample size. 
This suggests that the accuracy can be ``smoothed'' by allowing the test sample to be drawn with replacement. 
An algorithm that samples test sets with replacement is the \emph{leave-one-out bootstrap estimator},  and its derivatives, such as the \emph{0.632 bootstrap}, and \emph{0.632+ bootstrap} \citep[Sec 7.11]{hastie_elements_2003}.
\begin{definition}[bLOO]
	\label{def:bloo}
	The \emph{leave-one-out bootstrap} estimate, bLOO, is the average accuracy of the holdout observations, over all bootstrap samples. 
	Denote by $\data^b$, a bootstrap sample $b$ of size $n$, sampled with replacement from $\data$. 
	Also denote by $C^{(i)}$ the index set of bootstrap samples not containing observation $i$.
	The leave-one-out bootstrap estimate, $\accEstim_{\algo}^{bLOO}$,  is defined as:
	\begin{align}
	\accEstim_{\algo}^{bLOO}:= \frac 1n \sum_{i=1}^{n} \frac{1}{|C^{(i)}|} \sum_{b \in C^{(i)}} \indicator{\hypFun{\data^b}{x_i}=y_i}.
	\end{align}
	An equivalent formulation, which stresses the Bootstrap nature of the algorithm is the following. 
	Denoting by $S^{(b)}$ the indexes of observations that are \emph{not} in the bootstrap sample $b$ and are not empty, 
	\begin{align}
	\accEstim_{\algo}^{bLOO} = \frac 1B \sum_{b=1}^{B} \frac{1}{|S^{(b)}|} \sum_{i \in S^{(b)}} \indicator{\hypFun{\data^b}{x_i}=y_i}.
	\end{align}
\end{definition}

\begin{figure}[ht]
	\centering
	\includegraphics[width=0.5\linewidth]{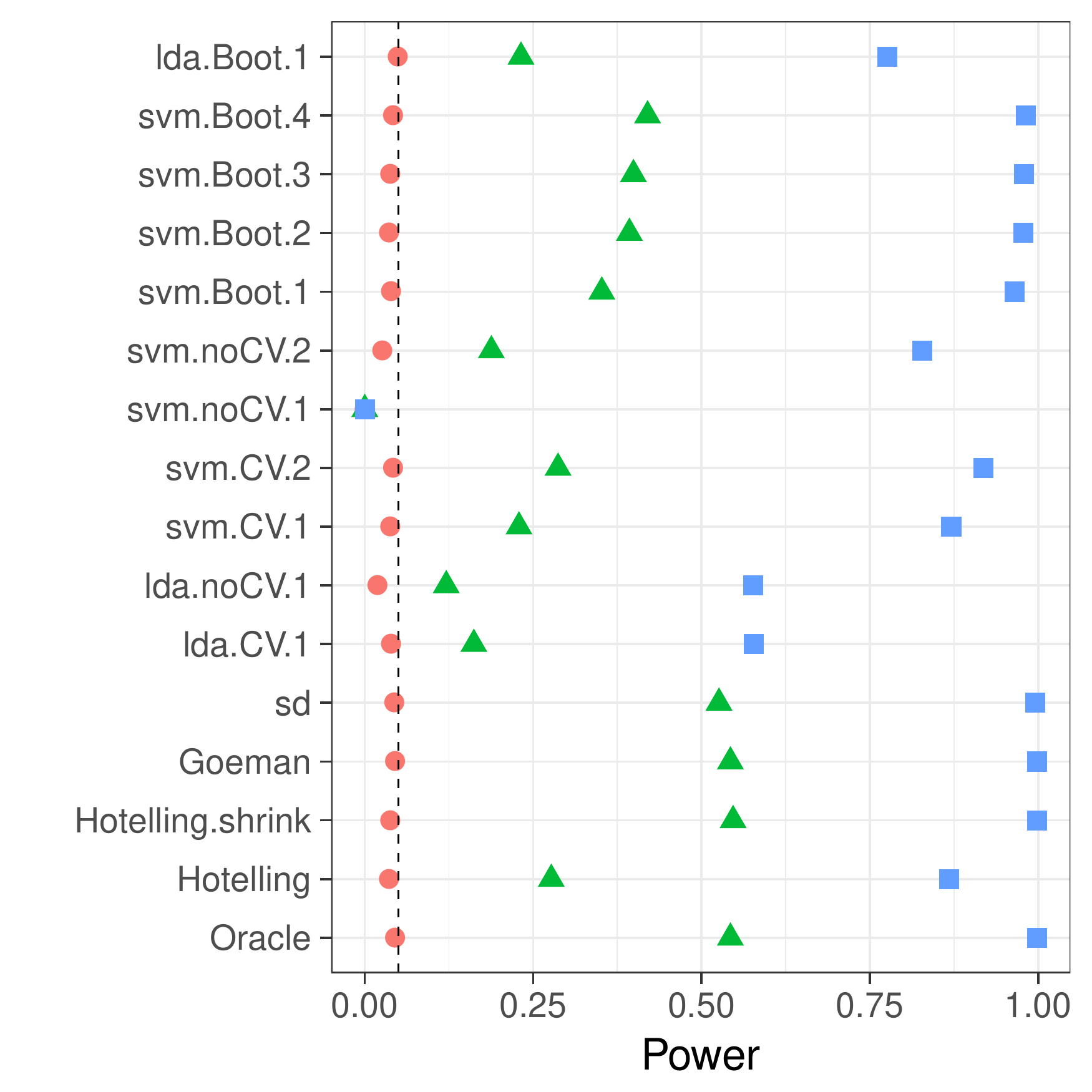}
	\caption{
		\textbf{Bootstrap.}
		The power of a permutation test with various test statistics. 
		The power on the $x$ axis. 
		Effects are color and shape coded. 
		The various statistics on the $y$ axis. 
		Their details are given in tables~\ref{tab:collected} and \ref{tab:collected_2}. 
		Effects vary over $0$ (red circle), $0.25$ (green triangle), and $0.5$ (blue square). 
		Simulation details in Appendix~\ref{sec:simulation_details}.
	} 
	\label{fig:bootstrap}
\end{figure}

Simulation results are reported in Figure~\ref{fig:bootstrap} with naming conventions in Table~\ref{tab:collected_2}.
As expected, selecting test sets with replacement does increase the power of accuracy tests, when compared to V-fold cross validation, but still falls short from the power of location tests. 
It can also be seen that power increases with the number of bootstrap replications, since more replications reduce the level of discretization.

\bigskip

\begin{tcolorbox}
	\centering
	\begin{tabular}{l|c|c|c|c}
		Name & Algorithm & Resampling & B  & Parameters\\ 
		\hline
		\hline
		LDA.Boot.1 & LDA & bLOO 	& $10$ &  -- \\ 
		SVM.Boot.1 & SVM & bLOO 	& $10$ & cost=10 \\ 
		SVM.Boot.2 & SVM & bLOO 	& $10$ & cost=0.1 \\ 
		SVM.Boot.3 & SVM & bLOO 	& $50$ & cost=10 \\ 
		SVM.Boot.4 & SVM & bLOO 	& $50$ & cost=0.1 \\ 
	\end{tabular} 
	\captionsetup{type=table}
	\caption{
		The same as Table~\ref{tab:collected} for bootstraped accuracy estimates. 
		bLOO is defined in~\ref{def:bloo}.
		$B$ denotes the number of Bootstrap samples.} 
	\label{tab:collected_2}
\end{tcolorbox}

\subsection{The Effect of High Dimension}
\label{sec:highdim}

Our setup of $n=40$ and $p=23$ is high dimensional in that $p/n$ is not too small. 
This surfaces finite samples effects, not manifested in classical $p/n \to 0$ asymptotic analysis.
Our best performing tests, \emph{sd}, \emph{Goeman}, and \emph{Hotelling.shrink}, alleviate the dimensionality of the problem by regularizing the estimation of $\Sigma$, thus reducing variance at the cost of some bias. 
It may thus be argued that the power advantages of the location tests are driven by the regularization of the covariance, and not the statistic itself.
We would thus augment the comparison with various covariance-regularized accuracy tests.
The $l_2$ regularization in our SVM accuracy test, already regularizes the covariance, but it is certainly not the only way to do so. 
We thus add some covariance-regularized accuracy tests such as a shrinkage based LDA \citep{pang_shrinkage-based_2009,ramey_high-dimensional_2016}, where similarly to \emph{Hotteling.shrink}, Tikhonov regularization of $\hat \Sigma$ is employed. 
We also try we try a diagonalized LDA\footnote{Known as \emph{Gaussian Na\"ive Bayes}.} \citep{dudoit_comparison_2002}, which regularizes $\hat{\Sigma}$ similarly to \emph{sd} and \emph{Goeman}.

\begin{figure}[ht]
	\centering
	\includegraphics[width=0.5\linewidth]{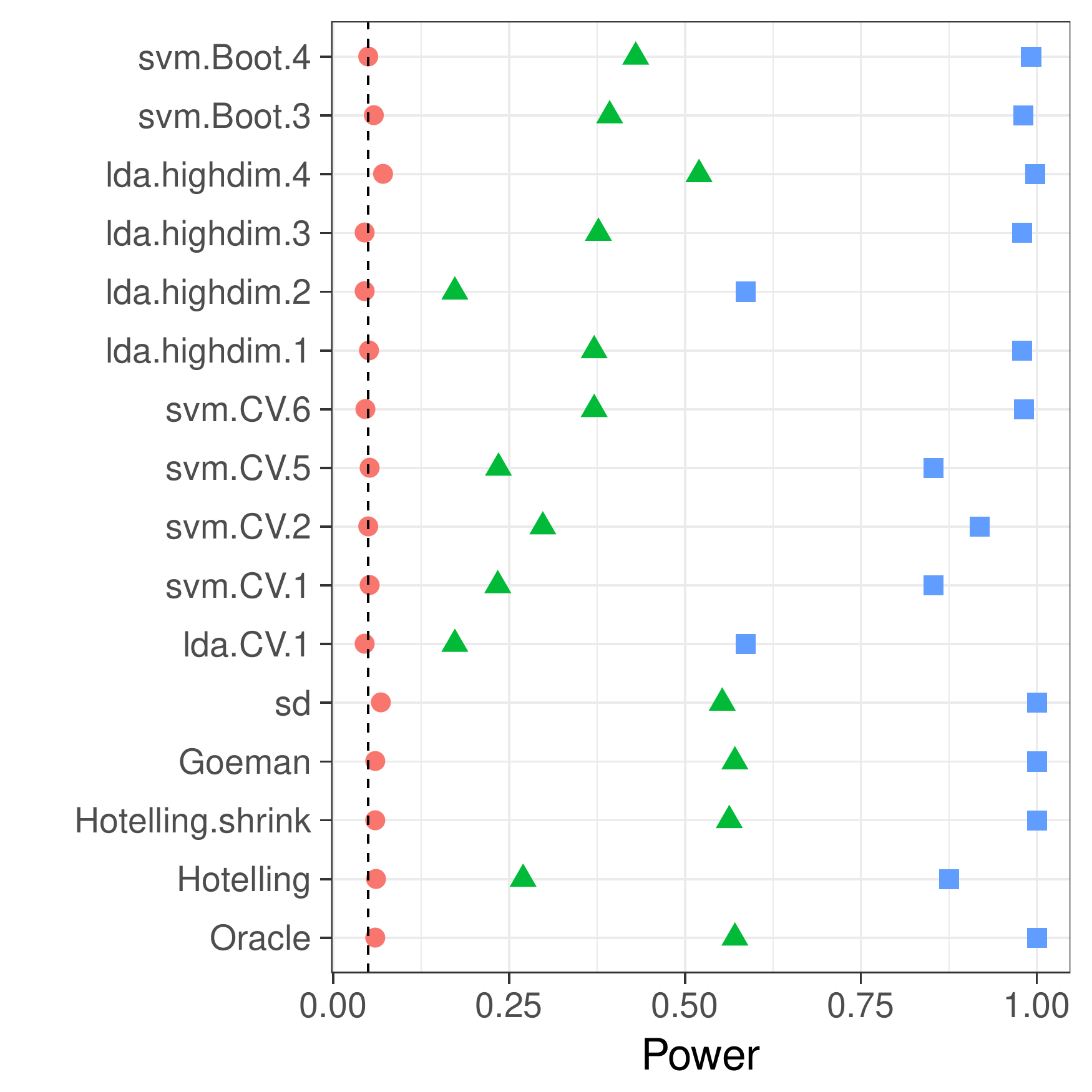}
	\caption{
		\textbf{HighDim Classifier.} 
		The power of a permutation test with various test statistics. 
		The power on the $x$ axis. 
		Effects are color and shape coded. 
		The various statistics on the $y$ axis. 
		Their details are given in tables~\ref{tab:collected} and \ref{tab:collected_3}. 
		Effects vary over $0$ (red circle), $0.25$ (green triangle), and $0.5$ (blue square). 
		Simulation details in Section~\ref{sec:simulation_details}.
	} 
	\label{fig:highdim}
\end{figure}

Simulation results are reported in Figure~\ref{fig:highdim} with naming conventions in Table~\ref{tab:collected_3}.
The proper regularization of the covariance of a classifier, just like a location test, can improve power. 
See, for instance, \emph{svm.CV.6} which is clearly the best regularized SVM for testing. 
Replacing the V-fold  with a bootstrap allows us to further increase the power, as done with \emph{lda.highdim.4}.
Even so, the out-of-the-box location tests outperform the accuracy tests.

\bigskip

\begin{tcolorbox}
	\centering
	\begin{tabular}{l|c|c|c}
		Name & Algorithm & Resampling &  Parameters\\ 
		\hline
		\hline
		svm.CV.5 & SVM & V-fold & cost=100 \\ 
		svm.CV.6 & SVM & V-fold & cost=0.01\\ 
		lda.highdim.1 & LDA & V-fold & -- \\ 
		lda.highdim.2 & LDA & V-fold & -- \\ 
		lda.highdim.3 & LDA & V-fold & -- \\ 
		lda.highdim.4 & LDA & bLOO 	 & B=50 \\ 
	\end{tabular} 
	\captionsetup{type=table}
	\caption{
		The same as Table~\ref{tab:collected} for regularized (high dimensional) predictors. 
		\emph{svm.CV.5} and \emph{svm.CV.6} are $l_2$ regularized SVM, with varying regularization penalty.
		\emph{lda.highdim.1} is the Diagonal Linear Discriminant Analysis of \cite{dudoit_comparison_2002}.
		\emph{lda.highdim.2} is the High-Dimensional Regularized Discriminant Analysis of \cite{ramey_high-dimensional_2016}.
		\emph{lda.highdim.3} is the Shrinkage-based Diagonal Linear Discriminant Analysis of \cite{pang_shrinkage-based_2009}.
		\emph{lda.highdim.4} is the same with bLOO.
	} 
	\label{tab:collected_3}
\end{tcolorbox}

\section{Neuroimaging Example}
\label{sec:example}

\begin{figure}[th]
\centering
\includegraphics[width=0.5\linewidth]{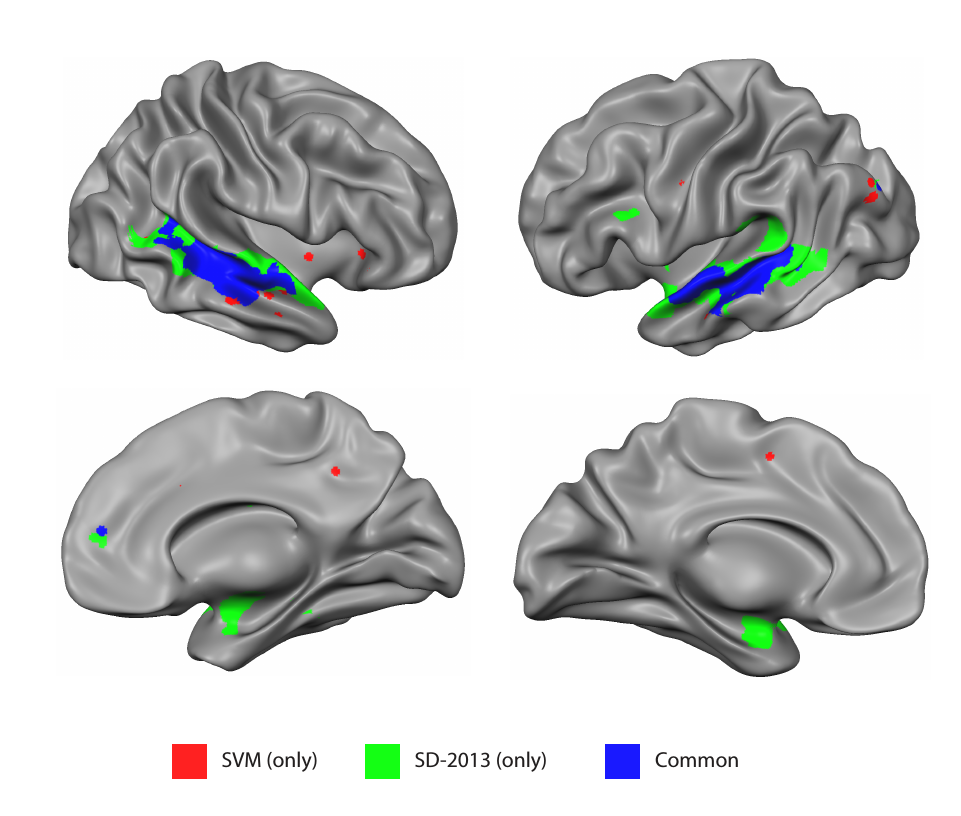}
\caption{\footnotesize
Brain regions encoding information discriminating between vocal and non-vocal stimuli.
Map reports the centers of $27$-voxel sized spherical regions, as discovered by an accuracy test (\emph{svm.CV.1}), and a location test (\emph{sd}). 
\emph{svm.CV.1} was computed using $5$-fold cross validation, and a cost parameter of $1$. 
Region-wise significance was determined using the permutation scheme of \cite{stelzer_statistical_2013}, followed by region-wise $FDR \leq 0.05$ control using the Benjamini-Hochberg procedure \citep{benjamini_controlling_1995}.
Number of permutations equals $400$.
The location test detect $1,232$ regions, and the accuracy test $441$, $399$ of which are common to both.
For the details of the analysis see \cite{gilron_quantifying_2016}.  
  }
\label{fig:read_data}
\end{figure}

Figure~\ref{fig:read_data} is an application of both a location and an accuracy test to the neuroimaging data of \cite{pernet_human_2015}. 
The authors of \cite{pernet_human_2015} collected fMRI data while subjects were exposed to the sounds of human speech (vocal), and other non-vocal sounds. 
Each subject was exposed to $20$ sounds of each type, totaling in $n=40$ trials.
The study was rather large and consisted of about $200$ subjects.
The data was kindly made available by the authors at the OpenfMRI website\footnote{\url{https://openfmri.org/}}.

We perform group inference using within-subject permutations along the analysis pipeline of \cite{stelzer_statistical_2013}, which was also reported in \cite{gilron_quantifying_2016}. 
To demonstrate our point, we compare the \emph{sd} location test with the \emph{svm.CV.1} accuracy test. 

In agreement with our simulation results, the location test (\emph{sd}) discovers more brain regions of interest when compared to an accuracy test (\emph{svm.CV.1}).
The former discovers $1,232$ regions, while the latter only $441$, as depicted in Figure~\ref{fig:read_data}.
We emphasize that both test statistics were compared with the same permutation scheme, and the same error controls, so that any difference in detections is due to their different power.

\section{Discussion}
\label{sec:discussion}

We have set out to understand which of the tests is more powerful: accuracy tests or location tests. 
Our current observation is that accuracy tests are never optimal.
There is always a multivariate test, possibly a location test, that dominates in power. 
Our advice to the practitioner is that location tests, in particular their regularized versions, are good performers in a wide range of simulation setups and empirically. 
They are also typically easier to implement, and faster to run, since no resampling is required. 
Their high-dimensional versions, such as \cite{schafer_shrinkage_2005}, \cite{goeman2006testing}, and \cite{srivastava_multivariate_2007}, are particularly well suited for empirical problems such as neuroimaging and genetics.

\subsection{Where do Accuracy Tests Lose Power?}
The low power of the accuracy tests compared to location tests can be attributed to the following causes: \newline
(a) \textbf{Discretization}: 
The discrete nature of accuracy test statistics. 
The degree of discretization is governed by the sample size. 
For this reason, an asymptotic analysis such as \cite{ramdas_classification_2016}, or \cite{golland_permutation_2005}, will not capture power loss due to discretization\footnote{This actually holds for all power analyses relying on a \emph{contiguity} argument \cite[Ch.6]{vaart_asymptotic_1998}.}.
An asymptotic analysis may suggest resubstitution accuracy estimates are good test statistics, while they suffer from very low finite-sample power. 
The canonical remedy for ties--- random tie breaking --- showed only a minor improvement (Sec.~\ref{sec:ties}). \newline
(b) \textbf{Shift Alternatives}: 
We focused on shift alternative so that location tests are expectedly superior via an NPL type argument.\newline
(c) \textbf{Inefficient} use of the data when validating with a holdout set. \newline
(d) Inappropriate \textbf{regularization} in high SNR regimes: testing requires less regularization than predicting. \newline

Given the above reasons and based on our professional experience, we dare argue that an accuracy test will rarely have more power than a high-dimensional location test.

\subsection{Interpretation}
Multivariate tests, and location tests in particular, are easier to interpret. 
To do so we typically use a NPL type argument, and think:
What type of signal is a test sensitive to?
What is the direction of the effect? etc.
Accuracy tests are seen as ``black boxes'', even though they can be analyzed in the same way. 
\citet{gilron2017s} demonstrate that the type of signal captured by accuracy tests is less interpretable to neuroimaging practitioners than location tests. 

Some authors prefer accuracy tests because they can be seen as effect-size estimates, invariant to the sample size. 
This is true, but the multivariate-statistics literature provides many multivariate effect-size estimators, that generalize Cohen's d, and do not suffer from discretization like accuracy estimates. 
Examples can be found, for instance, in \cite{stevens2012applied} and references therein.

\subsection{Fixed SNR}
\label{sec:fix_snr}

For a fair comparison between simulations, in particular between those with different $\Sigma$, we needed to fix the difficulty of the problem.
We defined ``a fair comparison'' to be such that a maximal power test would have the same power, justifying our choice of fixing the Mahalanobis norm of $\mu$. 
Formally, in all our simulations we set $\Vert \mu \Vert_\Sigma^2=c^2 p$.

Our choice implies that the Euclidean norm of $\mu$ varies with the covariance, and with the direction of the signal.
An initial intuition may suggest that detecting signal in the low variance PCs is easier than in the high variance PCs. 
This is true when fixing $\Vert \mu \Vert_2$, but not when fixing $\Vert \mu \Vert_\Sigma$.

\begin{figure}[h]
	\centering
	\caption{Short memory, AR(1) correlation. $\Vert \mu \Vert_2$ fixed. }	
	\label{fig:dependence_4}	
	\begin{subfigure}[t]{.4\textwidth}
		\centering
		\includegraphics[width=1\linewidth]{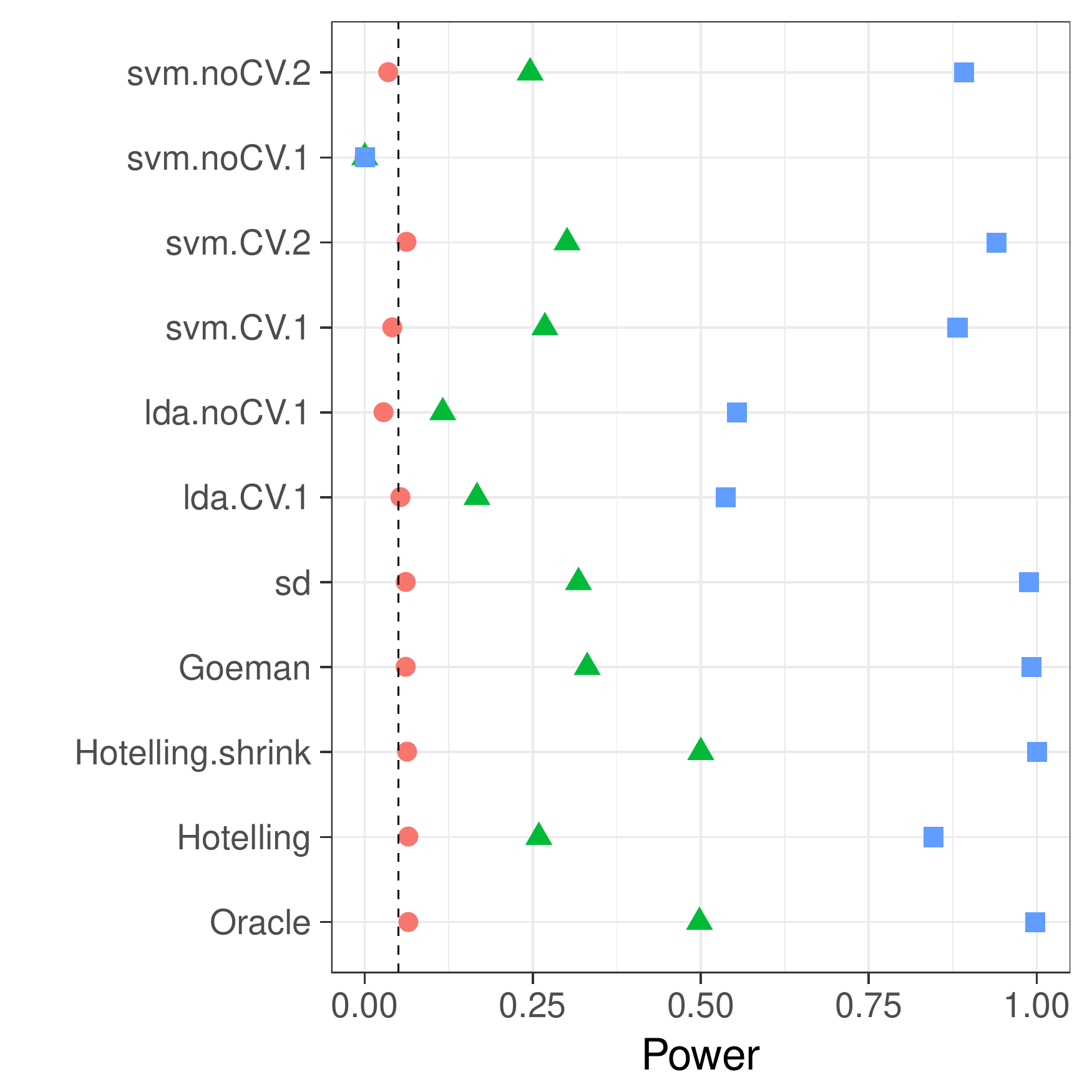}
		\caption{$\mu$ in PC7 of $\Sigma$.}  
		\label{fig:dependence_41}	
	\end{subfigure}
	\begin{subfigure}[t]{0.4\textwidth}
		\centering
		\includegraphics[width=1\linewidth]{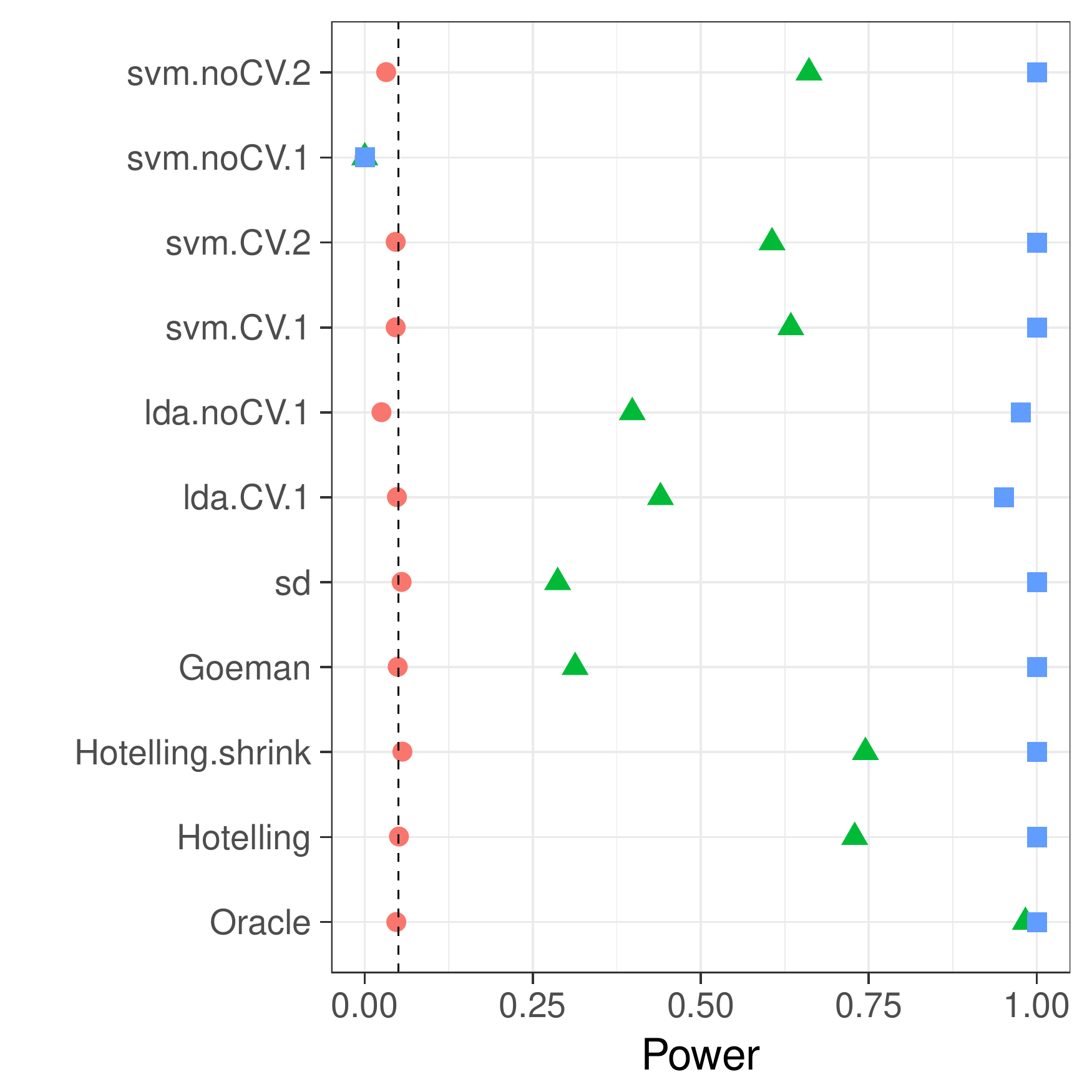}
		\caption{$\mu$ in PC15 of $\Sigma$.}  
		\label{fig:dependence_42}	
	\end{subfigure}

\end{figure}

For completeness, Figure~\ref{fig:dependence_4} reports the power analysis under $AR(1)$ correlations, but with $\Vert \mu \Vert_2$ fixed instead of $\Vert \mu \Vert_\Sigma$.
We compare the power of a shift in the direction of some high variance PC (Figure~\ref{fig:dependence_41}), versus a shift in the direction of a low variance PC (Figure~\ref{fig:dependence_42}).
The intuition that it is easier to detect signal in the low variance directions is confirmed. 
It is also consistent with Figure~\ref{fig:dependence_1}, in the following aspects: 
(i)\emph{Hotelling.shrink} is a good performed ``on average'', 
(ii) \emph{sd} and \emph{Goeman} have the best power to detect signal in the noisiest directions, but low power for signal in the noiseless directions.

\subsection{Detecting Signal in Different Directions}
Figures \ref{fig:dependence_1}, \ref{fig:dependence_2}, \ref{fig:dependence_3} and \ref{fig:dependence_4}, demonstrate that detecting signal in the direction of the high variance PCs is very different than detecting in the low variance PCs.
Why is that?

We attribute this phenomenon to regularization.
While the signal, $\mu$ varies in direction, the regularization of $\hat \Sigma$ does not. 
The various regularization methods deflate the high variance directions, thus, relatively inflate the low variance directions.
If the signal is in the low variance directions, the regularization may mask it. 
This is what we see in figures \ref{fig:dependence_12}, \ref{fig:dependence_22}, and \ref{fig:dependence_32}: the unregularized tests have more power than the regularized.

\subsection{Implications to Other Problems}

Our work studies signal detection in the two-group multivariate testing framework, i.e., MANOVA framework.
The same problem can be cast in the univariate generalized linear models framework, and in particular, as a Brenoulli Regression problem.
If any of the predictors, $x$, carries any signal, then $x|y=0$ has a different distribution than $x|y=1$.
This view is the one adopted \cite{goeman2006testing}.

Another related problem is that of multinomial-regression, i.e., multi-class classification.
We conjecture that power differences in favor of location tests versus accuracy tests will increase as the number of classes increases.

\subsection{Testing in Augmented Spaces}
It may be argued that only accuracy tests permits the separation between classes in augmented spaces, such as in \emph{reproducing kernel Hilbert spaces} (RKHS) by using non-linear predictors. 
This is a false argument--- accuracy tests do not have any more flexibility than location tests. 
Indeed, it is possible to test for location in the same space the classifier is learned. 
For independence tests  with kernels see for example \cite{szekely_brownian_2009} or \citet{gretton_kernel_2012-1}.

\subsection{A Good Accuracy Test}
Brain-computer interfaces and clinical diagnostics \citep[e.g.][]{olivetti_induction_2012,wager_fmri-based_2013} are examples where we want to know not only if information is encoded in a region, but rather, that a particular predictor can extract it. 
In these cases an accuracy test cannot be replaced by a location, or other, statistical test. 
For the cases an accuracy test cannot be replaced with other tests, we collect the following observations.

\paragraph{Sample size.} The conservativeness of accuracy tests, due to discretization, decrease with sample size. 

\paragraph{Regularize.}
Regularization proves crucial to detection power in low SNR regimes, such as when $n$ is in the order of $p$, or under strong correlations.
We find that the Shrinkage-based Diagonal Linear Discriminant Analysis of \cite{pang_shrinkage-based_2009} is a particularly good performer, but more research is required on this matter. 
Particularly, in the possibility of regularizing in directions orthogonal to $\mu$.

\paragraph{Smooth accuracy.}
Smooth accuracy estimate by cross validating with replacement. 
The bLOO estimator, in particular, is preferable over V-fold.

\paragraph{Resubstitution accuracy in high SNR.} 
Resubstitution accuracy is useful in high SNR regimes, such as $n \gg p$, because it avoids cross validation without compromising power. 
In low SNR, the power loss is considerable. 
We attribute this to the compounding of discretization and concentration effects: the difference between the sampling distribution of the resubstitution accuracy is simply indistinguishable under the null and under the alternative. 
In high SNR, the concentration is less impactful, and the computational burden of cross validation can be avoided by using the resubstitution accuracy. \newline

\subsection{Related Literature}
We now review some related accuracy-testing literature, with an emphasis on neuroimaging applications.
\cite{ojala_permutation_2010} study the power of two accuracy tests differing in their permutation scheme:
One testing the ``no signal'' null hypothesis, and the other testing the ``independent features'' null hypothesis. 
They perform an asymptotic analysis, and a simulation study. 
They also apply various classifiers to various data sets. 
Their emphasis is the effect of the underlying classifier on the power, and the potential of the ``independent features'' test for feature selection.
This is a very different emphasis from our own.

\cite{olivetti_induction_2012} and \cite{olivetti_statistical_2014} looked into the problem of choosing a good accuracy test. 
They propose a new test they call an \emph{independence test}, and demonstrate by simulation that it has more power than other accuracy tests, and can deal with non-balanced data sets. 
We did not include this test in the battery we compared, but we note that the independence test of \cite{olivetti_induction_2012} relies on a discrete test statistic. 
It may thus be improved by regularizing and resampling with replacement.

\cite{schreiber2013statistical} used null simulations to study the statistical properties of linear SVM's for signal detection, and in particular, false positive rates. 
They did not study the matter of power.
They recommended to test the significance of accuracy estimates using permutation testing instead of parametric t-tests, or binomial tests.
They recommend so due to the correlations between data folds in V-fold CV.
The authors were also concerned with temporal correlations, which biases accuracy estimates even if cross validated. 
Bias in accuracy estimates is of great concern when studying a classifier, but it is of lesser concern when using the accuracy merely for localization. 
Their recommendations differ from ours: they recommend to ensure independent data foldings in V-fold CV, whereas we claim discretization is the real concern, and thus recommend bLOO.

\cite{golland_permutation_2003} and \cite{golland_permutation_2005} study accuracy tests using simulation, neuroimaging data, genetic data, and analytically.
The finite Vapnik–Chervonenkis dimension requirement \citep[Sec 4.3]{golland_permutation_2005} implies a the problem is low dimensional and prevents the permutation p-value from (asymptotically) concentrating near $1$. 
They find that the power increases with the size of the test set.
This is seen in Fig.4 of \citet{golland_permutation_2005}, where the size of the test-set, $K$, governs the discretization. 
We attribute this to the reduced discretization of the accuracy statistic.

\cite{golland_permutation_2005} simulate the power of accuracy tests by sampling from a Gaussian mixture family of models, and not from a location family as our own simulations. 
Under their model (with some abuse of notation)
\begin{align*}
\begin{split}
	(x_i|y_i=1) & \sim \pi \gauss{\mu_1,I}+ (1-\pi) \gauss{\mu_2,I}, \\
	(x_i|y_i=0) & \sim (1-\pi) \gauss{\mu_1,I}+ \pi \gauss{\mu_2,I}.
\end{split}
\end{align*}
Varying $\pi$ interpolates between the null distribution $(\pi=0.5)$ and a location shift model $(\pi=0)$. 
We now perform the same simulation as \cite{golland_permutation_2005}, and in the same dimensionality as our previous simulations.
We re-parameterize so that $\pi=0$ corresponds to the null model:
\begin{align}
\begin{split}
\label{eq:mixture_alternative}
	(x_i|y_i=1) & \sim (1/2-\pi) \gauss{\mu_1,I}+ (1/2+\pi) \gauss{\mu_2,I}, \\
	(x_i|y_i=0) & \sim (1/2+\pi) \gauss{\mu_1,I}+ (1/2-\pi) \gauss{\mu_2,I}.	
\end{split}
\end{align}

\begin{figure}[ht]
\centering
	  \includegraphics[width=0.5\linewidth]{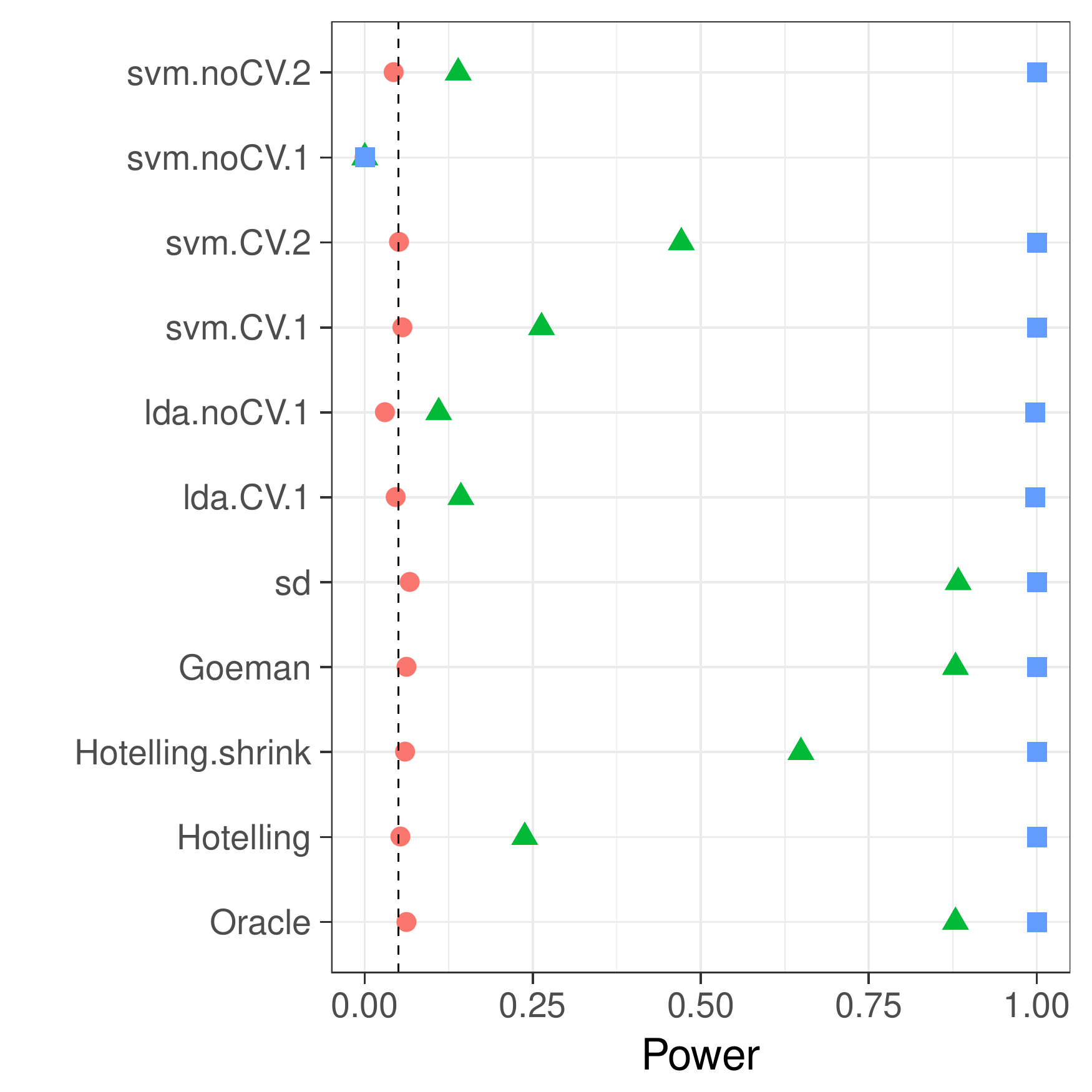}
	  \caption{\textbf{Mixture Alternatives.} $\x_i$ is distributed as in Eq.(\ref{eq:mixture_alternative}). 
	  	$\mu$ is a $p$-vector with $3/\sqrt{p}$ in all coordinates.
	  The effect, $\pi$, is color and shape coded and varies over $0$ (red circle), $1/4$ (green triangle) and $1/2$ (blue square). }
	\label{fig:file12}
\end{figure}

From Figure~\ref{fig:file12}, we see that also for the mixture class of \cite{golland_permutation_2005} locations tests are to be preferred over accuracy tests.

\subsection{Epilogue}
Given all the above, we find the popularity of accuracy tests for signal detection quite puzzling. 
We believe this is due to a reversal of the inference cascade. 
Researchers first fit a classifier, and then ask if the classes are any different.
Were they to start by asking if classes are any different, and only then try to classify, then location tests would naturally arise as the preferred method. 
As put by \cite{ramdas_classification_2016}:
\begin{quote}
The recent popularity of machine learning has resulted in the extensive teaching and use
of prediction in theoretical and applied communities and the relative lack of awareness or
popularity of the topic of Neyman-Pearson style hypothesis testing in the computer science
and related ``data science'' communities.
\end{quote}

\section*{Acknowledgments}
JDR was supported by the ISF 900/60 research grant. 
JDR also wishes to thank, Jesse B.A. Hemerik, Yakir Brechenko, Omer Shamir, Joshua Vogelstein, Gilles Blanchard, and Jason Stein for their valuable inputs.

\newpage
\bibliographystyle{abbrvnat}
\bibliography{Permuting_Accuracy.bib}

\end{document}